\documentclass[letterpaper,aps,prl,twocolumn,floatfix,amsfonts,amssymb,notitlepage,superscriptaddress,longbibliography]{revtex4-2}
\pdfpageattr {/Group << /S /Transparency /I true /CS /DeviceRGB>>}

\usepackage{amsmath}
\usepackage{amsthm}
\usepackage{amssymb}
\usepackage{graphicx}
\usepackage{tabularx}
\usepackage{color}
\usepackage{hyperref}
\usepackage{xcolor}
\usepackage[normalem]{ulem}

\newcommand{\figone}[0]{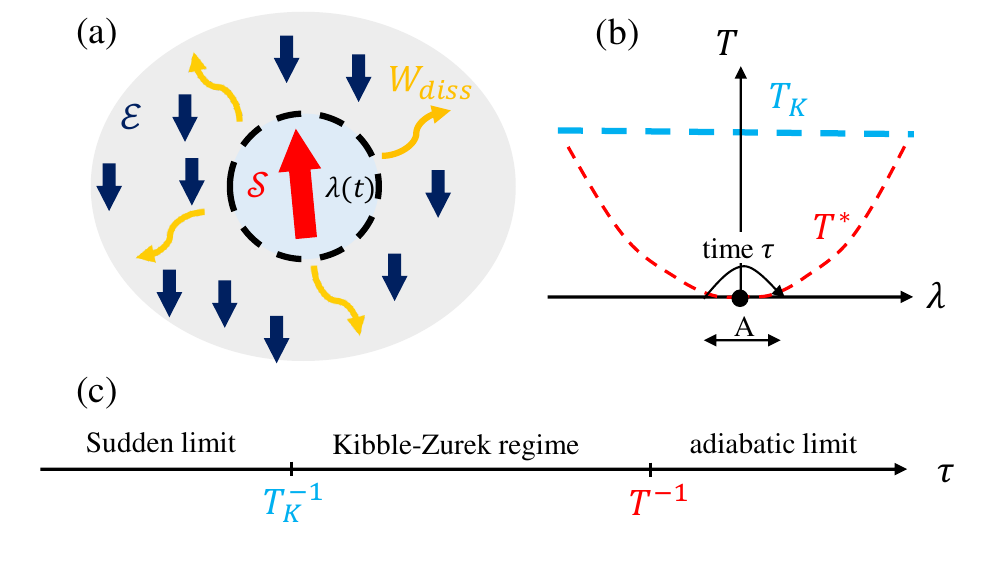}
\newcommand{\figtwo}[0]{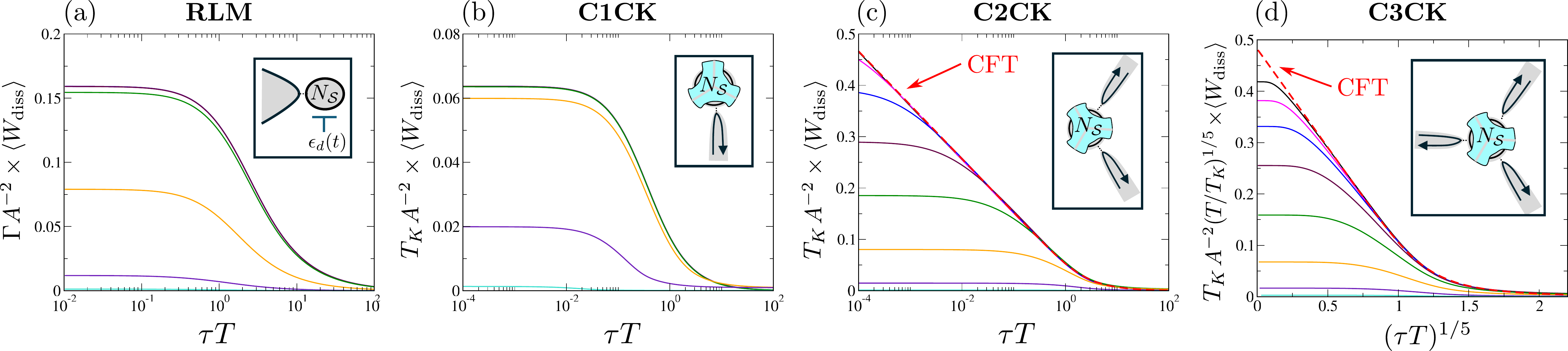}
\newcommand{\figthree}[0]{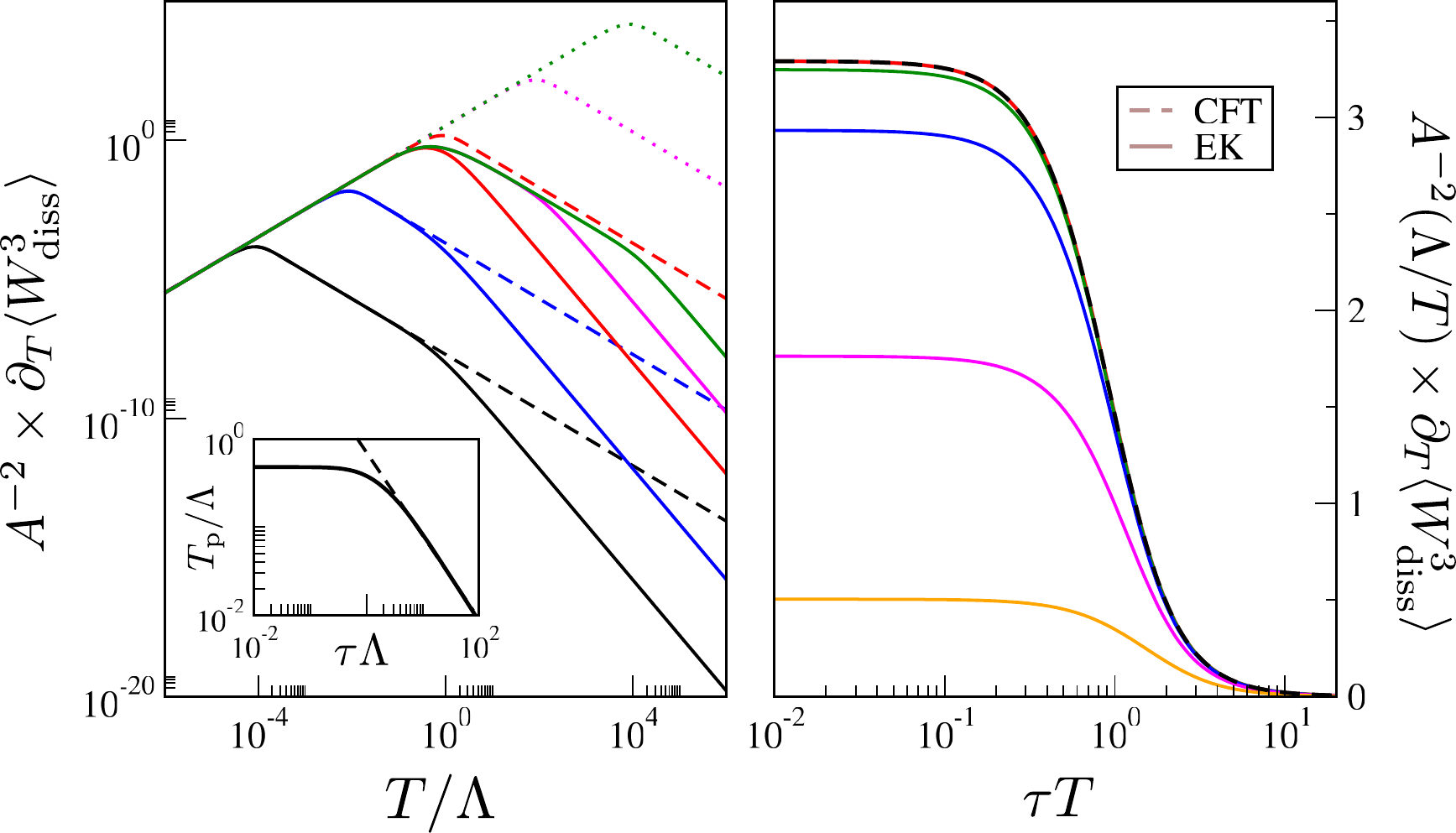}
\newcommand{\figfour}[0]{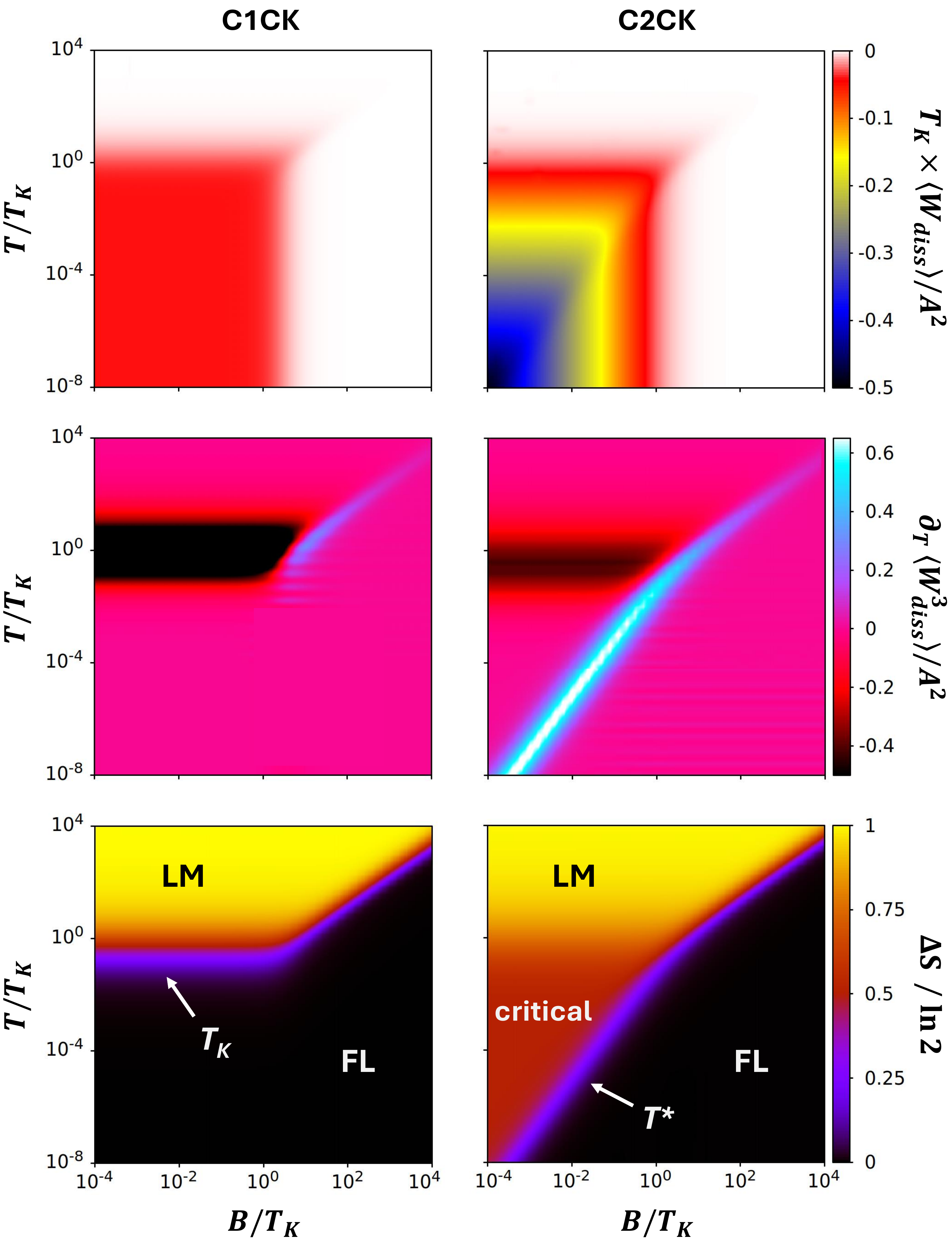}

\newcommand{\Fig}[1] {{\textcolor{black}{Fig.~}}~\!\!\ref{#1}}
\newcommand{\be}{\begin{equation}}
\newcommand{\ee}{\end{equation}}
\newcommand{\bea}{\begin{eqnarray}}
\newcommand{\eea}{\end{eqnarray}}

\newcommand{\Tr}{\text{Tr}}

\def\bs#1\es{\begin{split}#1\end{split}}	\def\bal#1\eal{\begin{align}#1\end{align}}

\begin{document}

\title{Quantum work statistics across a critical point:\\
full crossover from sudden quench to the adiabatic limit}

 \author{Zhanyu Ma}
\affiliation{Raymond and Beverly Sackler School of Physics and Astronomy, Tel Aviv University, Tel Aviv 69978, Israel} 
\author{Andrew K. Mitchell}
\affiliation{School of Physics, University College Dublin, Belfield, Dublin 4, Ireland} 
\affiliation{Centre for Quantum Engineering, Science, and Technology, University College Dublin, Dublin, Ireland}
\author{Eran Sela}
\affiliation{Raymond and Beverly Sackler School of Physics and Astronomy, Tel Aviv University, Tel Aviv 69978, Israel}

%###################
%###################

\begin{abstract}
\noindent When an external parameter drives a system across a quantum phase transition at a finite rate, work is performed on the system and entropy is dissipated, due to creation of excitations via the Kibble–Zurek mechanism.
Although both the adiabatic and sudden-quench limits have been studied in detail, the quantum work statistics along the crossover connecting these limits has largely been an open question. 
Here we obtain exact scaling functions for the work statistics along the full crossover from adiabatic to sudden-quench limits for critical quantum impurity problems, by combining linear response theory, conformal field theory, and the numerical renormalization group. These predictions can be tested in charge-multichannel Kondo quantum dot devices, where the dissipated work corresponds to the creation of nontrivial excitations such as Majorana fermions or Fibonacci anyons. 
\end{abstract}
\maketitle

%###################
%###################

Universality emerges near classical or quantum phase transitions, and determines equilibrium as well as non-equilibrium properties \cite{sachdev}. 
For example, when a system is driven in a finite amount of time across a quantum critical point (QCP), excitations are created at a rate dictated by Kibble-Zurek (KZ) scaling~\cite{kibble1976topology,zurek1985cosmological}. Universal non-equilibrium behavior is also found in the sudden-quench (impulse) limit, when a system is subjected to an abrupt change in parameters~\cite{calabrese2016quantum}.

Such non-equilibrium processes can be characterized by the work done on the system \cite{silva2008statistics,fei2020work,rossini2021coherent}.
The full work distribution function (WDF) contains rich information about the non-equilibrium state of a system, including quantum coherence effects, and satisfies fluctuation theorems such as Jarzynski’s equality~\cite{jarzynski1997nonequilibrium}.
Recently, the fundamental properties of the WDF for quantum many-body systems have been the subject of intense study. In particular, the irreversible work in the sudden limit is known to diverge at a QCP like a susceptibility~\cite{PhysRevLett.109.160601,PhysRevX.4.031029,goold2018role,PhysRevE.89.062103,PhysRevB.90.094304}. 
Away from the sudden limit, general statements about entropy production can be made~\cite{deffner2017kibble}, and in the adiabatic limit of slow driving, the full WDF can be calculated because the system remains close to equilibrium~\cite{PhysRevResearch.2.023377}. 
In this case, quantum coherence is shown to induce non-Gaussianity in the WDF and the implications of this for Landauer information erasure have been explored~\cite{PhysRevLett.128.010602,PhysRevLett.125.160602}.
Interestingly, only the higher moments of the WDF display quantum coherent effects. Non-Gaussian WDFs have also been observed for finite systems away from the adiabatic limit~\cite{zawadzki2020work,zawadzki2023non}. 
However, exact results for the WDF of true quantum many-body systems along the full crossover from the sudden-quench limit, through the KZ regime, to the adiabatic limit have proved elusive. This is because strongly-correlated quantum critical systems, in the thermodynamic limit and out of equilibrium, are notoriously difficult to treat -- either analytically or numerically.

One promising route forward is to consider the linear response (LR) regime, in which a \textit{weak} perturbation is applied to a critical system. This has the advantage that non-equilibrium properties can be related to equilibrium correlation functions, with no restriction on the time $\tau$ over which the driving is applied -- from sudden ($\tau\to 0$) all the way through to adiabatic ($\tau\to \infty$). The system-environment coupling can also be strong.

\begin{figure}
\centering
\includegraphics[width=\columnwidth]{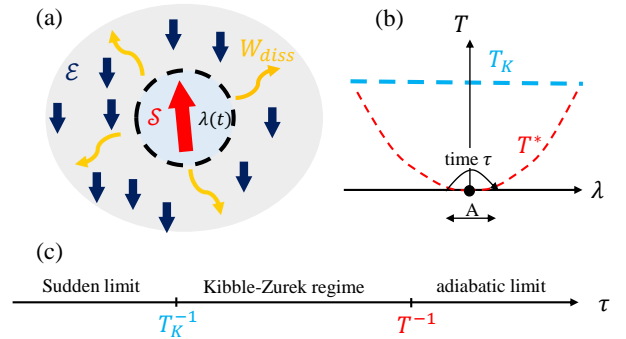}
\caption{
(a) System $\mathcal{S}$ and environment $\mathcal{E}$ are coupled, but share a conserved charge. In a quantum dot setup, the dot spin (red) can be flipped with a compensating spin-flip of an electron in the lead (blue). 
We consider the dissipated work, due to a weak perturbation $\lambda(t)$ such as a magnetic field, ramped in a finite time $\tau$. (b) Multichannel Kondo systems exhibit quantum critical physics and universality in the intermediate regime $T^*<T<T_K$, where $T_K$ is the Kondo temperature and $T^*\to 0$ at the QCP. %The perturbation $\lambda(t)$ drives the system across the QCP. 
(c) We obtain the full, universal crossover from sudden-quench to the adiabatic limit, including the intermediate KZ regime. 
}
\label{fig:schematic}
\end{figure}

Remarkable experiments with quantum dot (QD) devices ~\cite{saira2012test,koski2013distribution,hofmann2017heat,hofmann2016equilibrium,barker2022experimental,PhysRevLett.131.220405} have demonstrated that the WDF can be extracted in the classical regime by driving the QD gate voltage while performing a weak continuous measurement of the QD charge. However, the WDF has not been measured in the quantum regime of such systems. In particular, the WDF across a quantum phase transition in the KZ regime has not yet been considered. 
QD devices may be a uniquely appealing platform to study critical quantum work statistics, since generalized quantum impurity models that support nontrivial QCPs can be realized and probed experimentally \cite{potok2007observation,*keller2015universal,Mebrahtu_2012,Mebrahtu_2013,iftikhar2015two,*mitchell2016universality,iftikhar2018tunable,han2022fractional,pouse2023quantum,*karki2023z}. Indeed, the required gate voltage control and measurement of the QD charge is possible in these kinds of system \cite{piquard2023observing,child2022entropy}. 

Motivated by this, here we develop a theory of the quantum work statistics of boundary critical models, 
exploiting universality of the QCP to obtain exact results in the case of a weak perturbation applied in finite time. We apply this to Kondo models describing recent QD experiments, 
and confirm our predictions using numerical renormalization group (NRG) calculations \cite{wilson1975renormalization,*bulla2008numerical,weichselbaum2007sum,mitchell2014generalized,*stadler2016interleaved,rigo2022automatic}.

%%%%%%%%%%%%%%%%%%%%%%%%%%%

\emph{Setup.--} We consider a generic open quantum system setup, consisting of a system $\mathcal{S}$ coupled to an environment $\mathcal{E}$, see \Fig{fig:schematic}(a). The full, unperturbed Hamiltonian is  $\hat{H}_0=\hat{H}_{\mathcal{S}}+\hat{H}_{\mathcal{E}}+\hat{H}_{\mathcal{SE}}$ where $\hat{H}_{\mathcal{SE}}$ describes the coupling between the system and environment. Suppose that $\mathcal{S}$ and $\mathcal{E}$ share a conserved charge $\hat{N}=\hat{N}_{\mathcal{S}}+\hat{N}_\mathcal{E}$ such that $[\hat{H}_0,\hat{N}]=0$, although $\hat{N}_{\mathcal{S}}$ can of course fluctuate. We then apply a perturbation coupling locally to the charge of the system $\hat{H}(t)=\hat{H}_0+\lambda(t)\hat{N}_{\mathcal{S}}$, where the work parameter $\lambda(t)=A t/\tau$ is ramped up from $\lambda=0$ to $\lambda=A$ over a finite time duration $\tau$ by an external agent \cite{jarzynski2012equalities}. Our main focus will be the dissipated work $W_{\rm diss} = W-\Delta F$ irreversibly spent to drive the system out of equilibrium from an initial thermal state \cite{PhysRevX.4.031029,esposito2009nonequilibrium}. Here $W$ is the stochastic quantum work defined through a two-time projective energy measurement at the beginning and end of the driving \cite{talkner2007fluctuation}, and $\Delta F=F(\tau)-F(0)$ is the change in \textit{equilibrium} free energy at the initial temperature $T$.
The WDF provides a complete stochastic thermodynamic description of the process \cite{talkner2007fluctuation},
\be
P(W)=\sum_{m_{\tau},{n_{0}}} \langle n_0|\rho_0 |n_0  \rangle \: p_{{m_{\tau}}|{n_{0}}} \:  \delta(W-E_{m_{\tau}}+E_{n_{0}}) \;.
\ee
Here $\rho_0$ is the initial thermal state and $p_{{m_{\tau}}|{n_{0}}} =|\langle m_{\tau} |U_{\tau}|n_0 \rangle|^2$ are the conditional probabilities, where we have used the instantaneous spectral decomposition $\hat{H}(t)=\sum_n E_{n_t}|n_t\rangle\langle n_t|$. The final (non-equilibrium) state of the system $\rho_{\tau}^{\phantom{\dagger}}=U_{\tau}^{\phantom{\dagger}}\rho_0^{\phantom{\dagger}} U_{\tau}^{\dagger}$ is obtained from the time-evolution operator $U_{\tau}$. In the adiabatic limit $p_{m|n} =\delta_{mn}$ whereas in the sudden limit $p_{{m_{\tau}}|{n_{0}}} =|\langle m_{\tau} |n_0 \rangle|^2$. The behavior for finite-time driving is highly nontrivial. This standard definition of the quantum work satisfies the fluctuation theorems~\cite{talkner2007fluctuation} and is closely related to the irreversible entropy~\cite{verley2014work,landi2021irreversible}. For a discussion of alternative formulations, see e.g.~Refs.~\cite{gherardini2024quasiprobabilities,strasberg2021first}.

One can obtain the $n^{\rm th}$ moment of the WDF $\langle W^n \rangle = (-1)^n \frac{d^n}{du^n} h(u)|_{u=0}$ from the generating function $h(u)=\int dW P(W) e^{-u W}$, where \cite{talkner2007fluctuation}, 
\be
\label{eq:generating}
h(u)={\rm{Tr}} [ U_\tau^{\dagger} e^{-u (\hat{H}_0+A\hat{N}_{\mathcal{S}})} U_\tau e^{-(\beta-u)\hat{H}_0} ]/\mathcal{Z}\;.
\ee
Useful results in the sudden-quench limit $\tau\to 0$ follow immediately from the Zassenhaus formula~\cite{SM, PhysRevB.90.094304}\nocite{kubo1957statistical, filippone:tel-00908428}, 
\be
\label{eq:zass_sudden}
\langle W^n\rangle=A^n\langle \hat{N}_{\mathcal{S}}^n\rangle_0 + \delta \mathcal{Q}_n,
\ee
where the first term is a purely classical contribution, and 
$\delta \mathcal{Q}_n$ is a correction coming from quantum coherences. Interestingly, $\delta \mathcal{Q}_1=\delta \mathcal{Q}_2=0$ such that the first two moments do not contain information on the inherently quantum part of the work. For the third moment, \cite{SM} $\delta \mathcal{Q}_3=\frac{A^2}{2} \langle [\hat{N}_{\mathcal{S}},[\hat{H}_0,\hat{N}_{\mathcal{S}}]] \rangle_0= -\frac{A^2}{2}\Tr[\rho_0 \hat{H}_{\mathcal{SE}}]$. Higher moments involve more complicated nested commutators.

Such a splitting into classical and quantum contributions has also been discussed in other settings~\cite{PhysRevLett.125.160602,francica2019role,santos2019role,kiely2023entropy}.

%%%%%%%%%%%%%%%%%%%%%%%%%%%%
 
\textit{Work statistics for a noninteracting QD.--} 
As the simplest example we consider driving the potential $\epsilon_d(t)\equiv \lambda(t)$ of a spinless single level QD (the system $\mathcal{S}$) connected to a metallic lead (the environment $\mathcal{E}$). The QD-lead system is described by the resonant level model (RLM), $\hat{H}(t)=\epsilon_d(t)d^{\dagger}d + \sum_k \epsilon_k^{\phantom{\dagger}} c_k^{\dagger}c_k^{\phantom{\dagger}} + \hat{H}_{\mathcal{SE}}$ where the QD-lead coupling $\hat{H}_{\mathcal{SE}}=\sum_k (V_k^{\phantom{\dagger}} c_k^\dagger d + {\rm H.c.})$ allows the QD electron number $\hat{n}=d^{\dagger}d \equiv \hat{N}_{\mathcal{S}}$ to fluctuate. 

At high temperatures $T \gg \Gamma$, where $\Gamma=\pi \nu \sum_k |V_k|^2$ is the level width and $\nu$ is the density of lead states at the Fermi energy, the dynamics is described by a master equation. Denoting the probability that the level is empty or occupied by $p_0$ and $p_1$, we have $\dot{p}_1=\Gamma [p_0 f(\epsilon_d(t))-p_1 (1-f(\epsilon_d(t)))]$ where $f(\epsilon)=(1+e^{\epsilon/T})^{-1}$ is the Fermi distribution. Depending on the stochastic trajectories $n(t)$, which classically jump between 0 and 1, the work is obtained in experiments from $W=\int_0^\tau dt\: \frac{d\epsilon_d}{dt}n(t)$. Using this definition, and with real-time charge detection of the QD, the full WDF has been extracted in the classical regime in various experiments~\cite{saira2012test,koski2013distribution,hofmann2017heat,hofmann2016equilibrium,barker2022experimental,PhysRevLett.131.220405}.  As observed experimentally~\cite{hofmann2017heat}, in the sudden limit $\tau \Gamma \ll 1$ the WDF acquires the line shape of two sharp peaks, $P(W)=\langle \hat{n} \rangle_0 \delta(W-A)+(1-\langle \hat{n} \rangle_0)  \delta(W)$. The corresponding moments follow as $\langle W^n \rangle=A^n \langle \hat{n} \rangle_0$. These are precisely the \textit{classical} contributions to the work statistics in the sudden limit from Eq.~\ref{eq:zass_sudden} (i.e.~$\delta\mathcal{Q}_n=0$).

At lower temperatures we expect quantum corrections. Even though the RLM is a very simple model (non-critical, non-interacting), the QD and lead states do become entangled and the dynamics is non-Markovian. Computing the exact result for the third moment of the quantum WDF, we get a term beyond that captured by the master equation, $\delta\mathcal{Q}_3=-\frac{1}{2}A^2\Tr[\rho_0 \hat{H}_{\mathcal{SE}}]$. For the RLM, exact expressions for $\langle \hat{n}\rangle_0$ and $\langle \hat{H}_{\mathcal{SE}}\rangle_0$ can be easily obtained using Green's function methods \cite{SM}.

%%%%%%%%%%%%%%%%%%%%

\begin{figure*}[t]
\centering
\includegraphics[width=\textwidth]{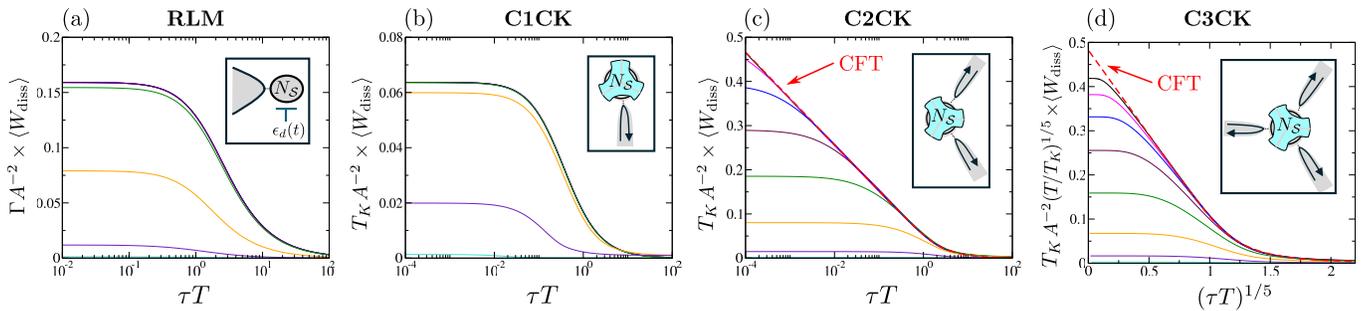}
\caption{Crossover in the dissipated work $\langle W_{\rm diss}\rangle$ due to a weak perturbation $\lambda(t)=At/\tau$ ramped over a finite time $\tau$, from sudden quench to adiabatic limits, for non-critical (a,b) and critical (c,d) quantum dot systems. (a) Spinless resonant level model describing a noninteracting quantum dot, subject to a ramp of the dot potential $\lambda(t)\equiv\epsilon_d(t)$, see inset.
(b,c,d)  Multichannel Kondo models with $M=1,2,3$ channels, respectively, subject to finite-time driving of the magnetic field $\lambda(t)\equiv B(t)$ (results obtained by NRG using $J=0.08D$). Such models describe charge-Kondo circuits with a metallic island coupled to $M$ leads, see insets. Dashed lines in (c,d) are the CFT scaling predictions derived from Eq.~\ref{eq:beta}. Shown for different temperatures $T/\Lambda=10^{n}$ with $n=-5 ... +2$ for black, magenta, blue, brown, green, orange, purple and cyan lines ($\Lambda=\Gamma$ for RLM and $T_K$ for CMCK).
} 
\label{fig:Wdiss}
\end{figure*}

%%%%%%%%%%%%%%%%%%%%

\textit{Linear response.--} Within LR (small $A$ limit) the WDF can be obtained from the relaxation function~\cite{guarnieri2023generalised},
\bea
\label{eq:relaxfnc}
\Psi_0(t)=\beta \int_0^\beta ds \: \langle \hat{N}_{\mathcal{S}}(-is) \hat{N}_{\mathcal{S}}(t) \rangle_0 -\beta^2 \langle \hat{N}_{\mathcal{S}}\rangle_0^2 \;,
\eea
where $\hat{N}_{\mathcal{S}}(t)=e^{i H_0 t}\hat{N}_{\mathcal{S}} e^{-i H_0 t}$. 
Determining the relaxation function is in general nontrivial. However, we find that its Fourier transform $\tilde{\Psi}_0(\omega)=%\tfrac{1}{2\pi}
\int dt \:\Psi_0(t) e^{
%-
i\omega t}$ is simply related \cite{SM} to the standard correlation function,
\be\label{eq:psi}
\tilde{\Psi}(\omega)= \frac{2}{\omega T} {\rm Im} \chi(\omega)+\frac{c(T)}{T}2\pi\delta(\omega),
\ee
where $\chi(\omega)\equiv \langle\langle \hat{N}_{\mathcal{S}};\hat{N}_{\mathcal{S}}\rangle\rangle_{\omega}$ is itself the Fourier transform of the retarded function $\chi(t)=i\theta(t){\rm Tr}\{\rho_0 [\hat{N}_{\mathcal{S}}(t),\hat{N}_{\mathcal{S}}]\}$. 
We find that $c(T)=0$ when the environment has a continuous gapless spectrum -- as for all systems we consider. 
In this case, cumulants of the WDF can be expressed as,
\begin{equation}\label{eq:cumulants}
    \frac{\kappa^n_W}{A^2} = 
    \begin{cases}
         \int \frac{d\omega}{2\pi}\: \omega^{n-2}\: {\rm sinc}^2(\tfrac{\omega\tau}{2})\: {\rm Im}\chi(\omega) & \text{: $n$ odd}\\
        \int \frac{d\omega}{2\pi}\: \omega^{n-2} \:{\rm sinc}^2(\tfrac{\omega\tau}{2}){\rm coth}(\tfrac{\omega}{2T})\: {\rm Im}\chi(\omega) & \text{: $n$ even}
    \end{cases}
\end{equation}
with ${\rm sinc}(x)\equiv\sin{x}/x$.
In particular, $\kappa_W^1\equiv \langle W_{\rm diss}\rangle$ is the expected dissipated work, and $\kappa_W^n=\langle (W_{\rm diss}-\langle W_{\rm diss}\rangle)^n\rangle$ are the central moments of $W_{\rm diss}$ %the dissipated work 
for $n=2,3$.

For the RLM describing a noninteracting QD, ${\rm Im}\chi(\omega)$ can be obtained exactly, and the resulting behavior of $\langle W_{\rm diss}\rangle$ is plotted in Fig.~\ref{fig:Wdiss}(a) along the full crossover from sudden to adiabatic at different temperatures.

Richer behavior of the work statistics can be expected in strongly-correlated systems, especially near a QCP.

%%%%%%%%%%%%%%%%

\textit{Charge-Kondo circuits.--} %Quantum 
Nanoelectronic circuits incorporating hybrid metal-semiconductor components perform as essentially perfect experimental quantum simulators of multichannel Kondo models~\cite{iftikhar2015two,*mitchell2016universality,iftikhar2018tunable,piquard2023observing,han2022fractional,pouse2023quantum,*karki2023z}. Near-degenerate macroscopic charge states $|N\rangle$ and $|N+1\rangle$ on a capacitive metallic island act as a charge pseudospin-$\tfrac{1}{2}$ degree of freedom $\hat{\boldsymbol{S}}_d$ (the system $\mathcal{S}$), that is flipped by tunneling at quantum point contacts to $M$ metallic leads (the environment $\mathcal{E}$). The charge $M$-channel Kondo (CMCK) model reads \cite{furusaki1995theory}
$\hat{H}_0=B\hat{S}_d^z + \sum_{m=1}^M \sum_{k \sigma} \epsilon_k^{\phantom{\dagger}} c^\dagger_{m k \sigma} c_{m k \sigma}^{\phantom{\dagger}} + \hat{H}_{\mathcal{SE}}$ where here the system-environment coupling is $\hat{H}_{\mathcal{SE}} = J(\hat{S}_d^+\hat{s}^-_c+{\rm H.c.})$, and $\hat{s}_c^-=\sum_{m}\sum_{kk'}c_{mk\downarrow}^{\dagger}c_{mk'\uparrow}^{\phantom{\dagger}}$. The effective magnetic field $B$ biases the island charge states, and is controlled in practice by a gate voltage. These models are highly non-Markovian open quantum systems in which strong, multipartite system-environment entanglement builds up at low temperatures $T\ll T_K$, where $T_K$ is the Kondo temperature \cite{affleck1993exact}. The $M=1$ version, %which is 
referred to as C1CK, is a non-critical system with a Fermi liquid ground state. All $M>1$ models, %which are 
referred to as C2CK, C3CK, etc., display boundary critical phenomena due to frustrated Kondo screening of the island charge pseudospin. At the QCP ($B=0$) the C2CK model is described by an effective Majorana RLM \cite{emery1992mapping,han2022fractional}, but finite $B$ induces a crossover scale $T^*\sim B^2/T_K$, see Fig.~\ref{fig:schematic}(b). For C3CK the island hosts a free Fibonacci anyon \cite{lopes2020anyons}.  
$M=1,2,3$ versions have been realized experimentally \cite{iftikhar2015two,*mitchell2016universality,iftikhar2018tunable,piquard2023observing}.

We analyze the quantum work statistics resulting from driving the effective field in finite time by varying the island gate voltage, where now $\hat{N}_{\mathcal{S}}=\hat{S}_d^z$ and $\lambda(t)=At/\tau$ as before. Our previous results (Eqs.~\ref{eq:zass_sudden} and \ref{eq:cumulants}) are general and carry over. All nontrivial information is therefore contained in the impurity dynamical spin susceptibility $\chi(\omega)=\langle\langle S_d^z;S_d^z\rangle\rangle_{\omega}$. In the sudden limit the first three moments can be obtained from the  
observables $\langle \hat{S}^z_d\rangle$ and $\langle \hat{S}_d^+\hat{s}^-_c + \hat{S}_d^-\hat{s}^+_c\rangle$. These must be computed in the full (lead-coupled but equilibrium) CMCK models.  %\zm{The LR condition is $A\ll 1/\chi$, with $\chi$ being the static charge susceptibility at $\lambda=0$, for RLM or CMCK.}

%%%%%%%%%%%%%%%%%%%%%

\begin{figure}
\centering
\includegraphics[width=0.96\columnwidth]{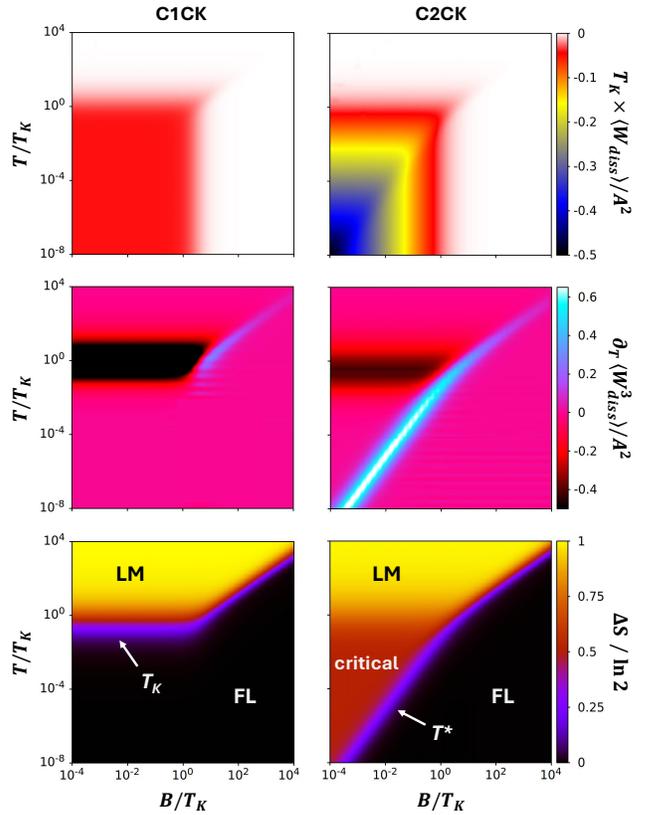}
\caption{Universal work statistics phase diagrams in the $(T/T_K,B/T_K)$ plane for the sudden-quench limit, comparing the C1CK model (non-critical, left panels) with the C2CK model (critical, right panels). The expected dissipated work $\langle W_{\rm diss}\rangle$ in the top row panels shows divergent behavior on approaching the QCP for $B \ll T_K$ and $T \ll T_K$ in the C2CK model (but saturation for C1CK) consistent with interpretation as a susceptibility. Middle row panels for $\partial_T\langle W_{\rm diss}^3\rangle$ show the appearance of the  critical scale $T^*$ in the C2CK model, reflected also in the thermodynamic entropy $\Delta S$ shown for reference in the bottom row panels (regimes labelled by their RG fixed points). Note that $\langle W_{\rm diss}^3\rangle$ and $\Delta S$ are connected by a Maxwell relation. Results obtained by NRG for $J=0.08D$.
}
\label{fig:sudden}
\end{figure}

\textit{C1CK model.--} Work statistics for the single-channel charge-Kondo model are obtained as above from ${\rm Im}\chi(\omega)$ computed from NRG~\cite{wilson1975renormalization,*bulla2008numerical,weichselbaum2007sum,mitchell2014generalized,*stadler2016interleaved,rigo2022automatic}. The system is non-critical and so we find that $\langle W_{\rm diss}\rangle$ shown in Fig.~\ref{fig:Wdiss}(b) behaves similarly to the RLM, despite being now a strongly interacting model -- except the C1CK scaling is in terms of the emergent Kondo temperature $T_K$, rather than the bare hybridization $\Gamma$ of the RLM. Further results in the sudden limit are discussed below in connection with Fig.~\ref{fig:sudden}.

%%%%%%%%%%%%%%%%%%%%

\textit{Dissipated work across a QCP.--} Figs.~\ref{fig:Wdiss}(c,d) for the C2CK and C3CK models show the evolution of the dissipated work along the crossover from sudden to adiabatic when driving across the multichannel Kondo critical point. From NRG results, we find universal scaling collapse in the KZ regime at intermediate driving times $T/T_K \ll \tau T \ll 1$, with $\langle W_{\rm diss}\rangle$ folding progressively onto a universal curve in both cases over a wider range of $\tau$ as the temperature $T$ is decreased. For $\tau T_K \ll 1$ the behavior departs again from KZ scaling and approaches the sudden limit. Note that signatures of the QCP show up in all moments of both the  dissipated work and the work itself. Exact results for the KZ and adiabatic scaling regimes (dashed lines) will be discussed in the following. 

%%%%%%%%%%%%%%%%%%%%

\textit{Universal results for boundary QCPs.--} For C2CK and C3CK models, the multichannel Kondo critical points are described \cite{affleck1993exact} by boundary conformal field theory (CFT). Importantly, we find the relaxation function is restricted to take a conformal-invariant form near the QCP,
\bea
\label{eq:scaling}
\Psi_0(t)=\beta \int_0^\beta ds\: \frac{1}{\Lambda^{2\Delta}} \bigg(\frac{\pi/\beta}{\sin \frac{\pi}{\beta}(s+it)} \bigg)^{2\Delta},
\eea
where $\Delta$ is the scaling dimension of the operator $\hat{N}_{\mathcal{S}}$ at the QCP (here $\hat{S}_d^z$). This result applies for \textit{any} QCP described by a boundary CFT, independently of its central charge.
This universal form applies for driving times $\tau$ much larger than the short time scale $\Lambda^{-1}$ associated with the QCP (for CMCK models the role of the cut-off $\Lambda$ is played by the Kondo temperature $T_K$) -- but it can be either large or small compared to the inverse temperature $\beta=1/T$. From Eq.~\ref{eq:scaling} one can write $\Psi_0(t)=\beta^{2-2\Delta} f_\Psi( t/\beta)$, where $f_\Psi(x)$ is some scaling function. This implies universal scaling of the work statistics for driving across the QCP, controlled only by the scaling dimension $\Delta$. For CMCK models where $\hat{N}_{\mathcal{S}}\equiv \hat{S}^z_d$, the scaling dimension of the impurity magnetization in the universal regime is $\Delta=\frac{2}{2+M}$ for $M \ge 2$. %Analytic expressions for $\Psi_0(t)$ can be obtained for any $\Delta$, as shown in the \textit{End Matter}.

Eqs.~\ref{eq:cumulants} and \ref{eq:scaling}
are general and allow the work statistics in LR to be obtained for any critical system described by a boundary CFT for any $\tau\gg \Lambda^{-1}$. Below we extract key results for the three scaling regimes identified in \Fig{fig:schematic}(c).
%Below we identify three scaling regimes, see \Fig{fig:schematic}(c). In particular, we find an intermediate KZ regime 
%$T_K^{-1} \ll \tau \ll T^{-1}$ between the sudden and adiabatic limits, 
%with unique anomalous scaling properties in terms of driving rate and temperature controlled by the CFT of the QCP.

%%%%%%%%%%%%%%%%%%%%%%%

%%%%%%%%%%%%%%%%%%%%%%%%

\noindent\textbf{Adiabatic limit:} 
For long ramps $\tau \gg T^{-1}$, Eq.~\ref{eq:cumulants} yields, 
\begin{eqnarray}
\label{eq:adia}
\langle W_{diss} \rangle_{\tau \to \infty}  =\tfrac{1}{2} T A^2 \tau^{-1} \tilde{\Psi}(\omega=0) \;.
\end{eqnarray}
Since the relaxation function does not depend on $\tau$, in the adiabatic limit the dissipated work decays as $1/\tau$ for any $\Delta$. Furthermore, the scaling form of $\Psi_0(t)$ implies specific temperature dependences. We find $\langle W_{diss} \rangle \sim T^{2\Delta-2}/\tau$ as well as $\kappa^2_W \sim  T^{2\Delta-1}/\tau$ and 
$\kappa^3_W \sim  T^{2\Delta-1}/\tau^2$.

%%%%%

\noindent\textbf{Sudden limit:} The field theory results do not strictly apply for $\tau\ll \Lambda^{-1}$, and the cut-off $\Lambda\equiv T_K$ in CMCK models is finite. To access the sudden limit we instead take $\tau\to 0$ in Eq.~\ref{eq:cumulants}, from which it follows that,
\be
\label{eq:kappa1dndl}
\langle W_{diss} \rangle_{\tau \to 0} =\tfrac{1}{2} T A^2 \Psi_0(t=0) \equiv - \frac{A^2}{2} \frac{d \langle \hat{N}_{\mathcal{S}} \rangle}{d \lambda}\Big|_{\lambda=0} \;,
\ee
where we have used the fact that the zero-time relaxation function is related to a susceptibility~\cite{SM}. This result also follows from Eq.~\ref{eq:zass_sudden} by writing $W_{\rm diss}=W-\Delta F$ and expanding $\Delta F$ to second order in $\lambda$, noting that $dF/d\lambda=\langle \hat{N}_{\mathcal{S}}\rangle$. 
%The latter result has been noted previously~\cite{PhysRevE.89.062103,goold2018role}. Consistent with Eq.~\ref{eq:zass_sudden}, for the third cumulant we find~\cite{SM},
Similarly for the third cumulant we find~\cite{SM},
\be
\kappa^3_W =A^2 \int \frac{d\omega}{2\pi} \: \frac{\omega^2}{2\beta }\tilde{\Psi}_0(\omega) = \delta \mathcal{Q}_3 \equiv -\frac{A^2}{2}\langle \hat{H}_{\mathcal{SE}}\rangle_0\;.
\label{eq:3nd_cu}
\ee
NRG results for C1CK and C2CK in the sudden limit, using $\hat{N}_{\mathcal{S}}\equiv \hat{S}^z_d$, are presented in Fig.~\ref{fig:sudden}. Top panels show the diverging dissipated work in the critical C2CK model for $T,B \ll T_K$, which follows from the log-diverging static magnetic susceptibility $d\langle S^z_d\rangle/dB$ near the QCP. By contrast, no divergence is seen in C1CK which has a saturating magnetic susceptibility and no QCP. 

Middle panels show $\partial_T \langle W^3_{\rm diss}\rangle$ obtained from $\kappa_W^3$ in Eq.~\ref{eq:3nd_cu}, which shows very clearly the vanishing $T^*$ scale in the vicinity of the QCP in the C2CK model. The C1CK model does not support this critical region, and is a Fermi liquid (FL) for all $T\ll \max(T_K,B)$. The structure of the phase diagram 
%in the $(T,B)$ plane 
for $\partial_T \langle W^3_{\rm diss}\rangle$ reflects that of the thermodynamic entropy change $\Delta S$ upon completely polarizing the impurity spin, as shown in the lower panels. This relation can be understood from Eq.~\ref{eq:3nd_cu} by applying the Maxwell relation $\partial_T \langle \hat{H}_{\mathcal{SE}}\rangle = -J\partial_J S$, 
which implies that a \textit{crossover} in the entropy is accompanied by a \textit{peak} in $|\partial_T \langle W^3_{\rm diss}\rangle|$, see \textit{End Matter} for details.

%%%%%%%%%%%%%%%%%%%%%%%%%

\noindent\textbf{Kibble-Zurek regime:} The most interesting physics arises in the KZ regime $T_K^{-1} \ll \tau \ll T^{-1}$ of the critical CMCK models. This is precisely where the CFT scaling results apply. As shown in the \textit{End Matter}, analysis of Eqs.~\ref{eq:cumulants} and Eq.~\ref{eq:scaling} leads to the following KZ scaling predictions (which hold in LR for finite-time driving across any boundary QCP). For the dissipated work we find,
\bea
\frac{(\Lambda/T)^{2\Delta-1}}{\Lambda^{-1}A^2} \:\langle W_{\rm diss} \rangle =
\begin{cases}
   \: -\ln{(\pi\tau T)}\phantom{\Big )} & \text{: $\Delta=1/2$}\\
   \: c_1-c_2\left(\pi\tau T \right)^{1-2\Delta}\phantom{\Big )}  & \text{: $\Delta < 1/2$}
\end{cases}\label{eq:Wdiss_kz}   
\eea
where $c_1$ and $c_2$ are positive constants. For C2CK and C3CK this behavior is confirmed in Fig.~\ref{fig:Wdiss}(c,d).   The full crossover curves for $\tau  \gg \tau_0$ can also be computed from Eq.~\ref{eq:scaling}. Exact results for all $\tau T$ can alternatively be obtained for the special case of C2CK using the Emery-Kivelson method \cite{emery1992mapping}, and are found to agree with the more general field theory predictions for $\Delta=1/2$ when $\tau \gg \tau_0$, see \textit{End Matter}.

%%%%%%%%%%%%%%%%%%%%%%%%%

\textit{Outlook.--} The nontrivial properties of the QCPs discussed here, which support boundary-localized Majorana fermions and Fibonacci anyons~\cite{emery1992mapping,han2022fractional,lopes2020anyons,PhysRevLett.129.227703}, are shown to have unique work distribution functions. %characterizing the non-equilibrium state when a perturbation is applied in a finite time.
%The entropy irreversibly produced in the KZ regime can be interpreted as being carried off by the previously localized fractional excitations, which are emitted into the bulk and lost when the system is driven.  
%
While charge-Kondo circuits realize these multichannel Kondo critical points~\cite{iftikhar2015two,*mitchell2016universality,iftikhar2018tunable,han2022fractional,pouse2023quantum,*karki2023z}, or alternatively spin-Kondo QCPs in semiconductor QDs~\cite{potok2007observation,*keller2015universal}, experimental challenges remain for extracting the WDF in the quantum regime~\cite{PhysRevB.110.115153}. 

Our results have implications for Landauer information erasure in quantum dot devices, extending the role of %The energy gap of an effective two-level open quantum system can be ramped up in practice by controlling the gate voltage or magnetic field on the dot. Although 
quantum coherence%effects have recently been discussed in erasure processes 
~\cite{PhysRevLett.128.010602,PhysRevLett.125.160602} %, our work allows the role of 
to the regime of strongly non-Markovian dynamics and many-body physics. % to be explored, beyond the physics captured by master equation methods. This is the experimentally-relevant regime for quantum dot devices at low temperatures, where ubiquitous  electron interaction effects produce strong system-environment entanglement. 
We also note a possible connection between  quantum work statistics and the quantum Fisher information due to a ramped LR perturbation \cite{mihailescu2024quantum}, since both quantities are controlled by the same relaxation function.

%Similar results are expected for bulk 1d critical systems where a CFT description applies. Indeed, the LR formalism can be used to study generic bulk systems, provided the relaxation function can be calculated. 
Notice that LR is valid in a limited validity regime, particularly in our models we require $A \ll \chi^{-1}$, with $\chi$ being the static charge susceptibility. While the latter diverges at $T=0$, a finite validity regime exists at finite temperature. The work statistics beyond LR remains  a largely open problem for %strongly-correlated 
many-body quantum systems. %Time-dependent NRG~\cite{anders2005real,nghiem2014time,*nghiem2014generalization} could however be used for quantum impurity models beyond LR.
\\
%%%%%%%%%%%%%%%%%%%%
%%%%%%%%%%%%%%%%%%%%

\begin{acknowledgments}
\noindent\textit{Note added in proof.--} While under review, we became aware of a related work, Ref.~\cite{PhysRevA.111.042207}.

\noindent{\textit{Acknowledgments.--}} We thank Mark Mitchison for pointing out the second term in Eq.~\ref{eq:psi} and Steve Campbell for useful discussions. ES and ZM gratefully acknowledge support from the European Research Council (ERC) under the European Union Horizon 2020 research and innovation programme under grant agreement No. 951541. AKM acknowledges support from Science Foundation Ireland through Grant 21/RP-2TF/10019.
\clearpage

%%%%%%%%%%%%%%%%%%%%
%%%%%%%%%%%%%%%%%%%%

\section{END MATTER}

\noindent\textit{Relaxation function as a dynamical susceptibility.--}
Eq.~\ref{eq:psi} is derived in the Supplementary Material~\cite{SM}. In general, it contains a delta-function contribution with the temperature-dependent coefficient $c(T)$. It is given by,
\be
c(T)=- {\rm Re}[\chi(\omega=0)] -\frac{d\langle N_{\mathcal{S}}\rangle}{d\lambda}\Big|_{\lambda=0}     \;.
\ee
For open quantum systems with an environment that has a continuous gapless spectrum, LR theory provides a connection between the dynamical susceptibility $\chi(\omega)\equiv 
i\int_{0}^{\infty} dt \: e^{-i\omega t}\:{\rm Tr}\{\rho_0 [\hat{N}_{\mathcal{S}}(t),\hat{N}_{\mathcal{S}}]\}$
and the static susceptibility $\frac{d\langle N_{\mathcal{S}}\rangle_0}{d\lambda}$ at a given temperature $T$. Specifically, ${\rm Re}[\chi(\omega=0)]=-\frac{d\langle N_{\mathcal{S}}\rangle_0}{d\lambda}$ and therefore $c(T)=0$. Thus Eq.~\ref{eq:cumulants} is exact for such systems in LR. For finite closed systems, or for systems coupled to an environment with a discrete spectrum, the Kubo formula for the static susceptibility no longer applies, and $c(T)$ is generically non-zero. In this case, the first two cumulants (only) in Eq.~\ref{eq:cumulants} pick up a correction \cite{SM}. We interpret this in terms of a heating of the system. This correction will typically vanish in the thermodynamic limit when the total heat capacity of the environment diverges.

%%%%%%%%

\noindent\textit{Derivation of Eq.~\ref{eq:Wdiss_kz}.--} The integral in Eq.~\ref{eq:scaling} can be performed explicitly for any $\Delta$ to obtain,
\be
\Psi_0(t)=\frac{\pi^{2\Delta-1} T^{2\Delta-2}}{\Lambda^{2\Delta}}e^{-i\pi\Delta}\sin{(\pi\Delta)}{\rm{B}}(-{\rm{csch}}^2 \pi t T ; \Delta, \frac{1}{2})
\label{eq:beta}
\ee
where ${\rm B}$ is the incomplete beta function. 
For $\Delta=\frac{1}{2}$,% (relevant to C2CK), Eq.~\ref{eq:beta} reduces to,
\begin{eqnarray}
\label{eq:2CK}
\Psi_0(t)=2(\Lambda T)^{-1} {\rm{arccoth}}(\cosh\pi t T).
\end{eqnarray}
The short-time ($\pi t T \ll 1$) behavior of $\Psi_0(t)$ is directly related to the KZ scaling of the work statistics. Expanding Eq.~\ref{eq:beta} as a series in $t$, we find,
\begin{multline}
\Lambda^2(\Lambda/T)^{2\Delta-2}\Psi_0(t)  =\\
\begin{cases}
   \: -2\ln(\pi t T) + \mathcal{O}(1) & \text{~: $\Delta=1/2$}\\
   \: \tilde{c}_1-\tilde{c}_2\left(\pi t T \right)^{1-2\Delta}+\mathcal{O}\left(t T\right)^{3-2\Delta}  & \text{~: $\Delta < 1/2$}
\end{cases}
\label{eq:Psi_scaling}
\end{multline}
where both $\tilde{c}_1\equiv\sin{(\pi\Delta)}\pi^{2\Delta-3/2}\Gamma(\frac{1}{2}-\Delta)\Gamma(\Delta)$ and $\tilde{c}_2\equiv2\sin{(\pi\Delta)}\pi^{2\Delta-1}(1-2\Delta)^{-1}$ are positive constants that depend on $\Delta$. Therefore as $t\to 0$, $\Psi_0(t)$ shows a logarithmic divergence for $\Delta=\frac{1}{2}$, but $\Psi_0(t)$ saturates to a finite value in a power-law fashion for $\Delta<\frac{1}{2}$.

For the dissipated work, Eq.~\ref{eq:cumulants} can be rewritten as $\langle W_{\rm diss} \rangle = TA^2\int_0^1 du (1-u)\Psi_0(\tau u)$. In the KZ regime $\tau T\ll 1$, and for $0<u<1$, $\Psi_0(\tau u)$ can be approximated by Eq.~\ref{eq:Psi_scaling}. Substituting it into the integral yields Eq.~\ref{eq:Wdiss_kz} with $c_1=\tilde{c}_1/2$ and $c_2=\tilde{c}_2/(2+3(1-2\Delta)+(1-2\Delta)^2)$.

Similarly, $\kappa_W^3$ can be expressed as $\kappa_W^3=TA^2\tau^{-2}(\Psi_0(0)-\Psi_0(\tau))$ from Eq.~\ref{eq:cumulants}. Combining this with Eq.~\ref{eq:Psi_scaling} yields in the KZ regime
\bea
  \frac{(\Lambda/T)^{2\Delta+1}}{\Lambda A^2}\:\kappa_W^3 = 
\begin{cases}
   \: 2\left(\tau T \right)^{-2}\ln{\left(\tau/\tau_0  \right)}\phantom{\Big )} & \text{~: $\Delta=1/2$}\\
   \: c_2^{\star}\left(\pi\tau T \right)^{-1-2\Delta}\phantom{\Big )}  & \text{~: $\Delta < 1/2$}
\end{cases} 
\label{eq:kappa3_kz}
\eea
where $c_2^*=\tilde{c}_2\pi^2$ and $\tau_0\sim T_K^{-1}$ is a UV cutoff, which is introduced because $\Psi_0(0)$ diverges for $\Delta=1/2$.%A full solution requires the knowledge of $\Psi_0(0)$, which is beyond the scope of the field-theoretical approach. 

%%%%%%%%%%%%%%

\noindent\textit{Dissipated work in the sudden limit as a susceptibility.--}
Eq.~\ref{eq:kappa1dndl} makes clear that the dissipated work as $\tau \to 0$ is related to the charge susceptibility of the system (or the static spin susceptibility in the Kondo language). This is vividly illustrated in Fig.~\ref{fig:Wdiss_sudden} where show $\langle W_{\rm diss}\rangle$ in the sudden limit upon reducing the temperature. For non-critical (Fermi liquid) systems such as RLM and C1CK, the dissipated work saturates at low $T$, whereas it diverges for the critical (non-Fermi liquid) C2CK and C3CK models. This is a characteristic hallmark of quenching a critical system. For C2CK $\langle W_{\rm diss}\rangle\sim \ln(T_K/T)$ whereas for C3CK $\langle W_{\rm diss}\rangle\sim (T_K/T)^{1/5}$.

\begin{figure}
    \centering
    \includegraphics[width=0.65\linewidth]{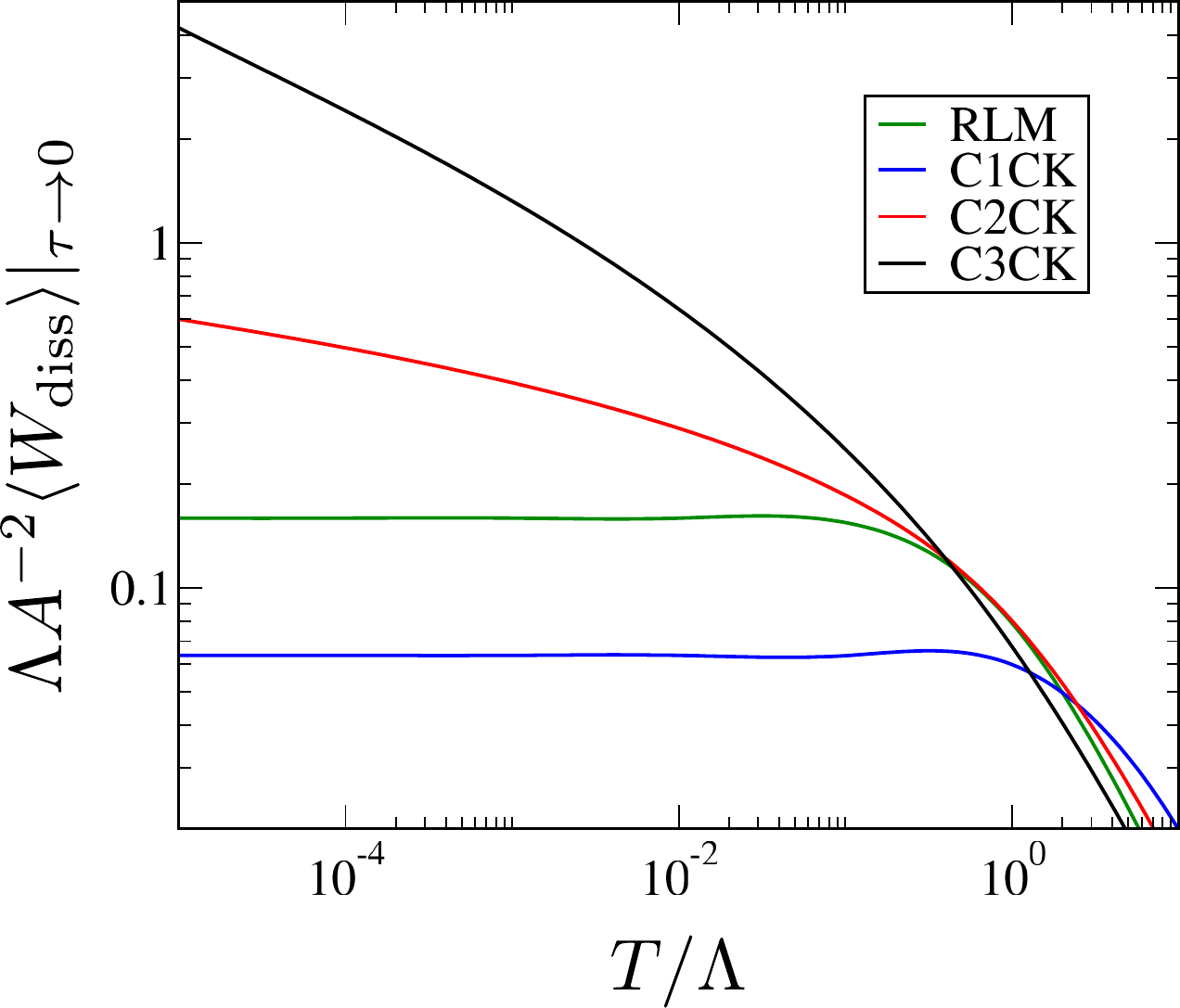}
    \caption{In the sudden-quench limit $\tau\to 0$ the dissipated work $\langle W_{\rm diss} \rangle$ in LR is proportional to a susceptibility, see Eq.~\ref{eq:kappa1dndl}. At low temperatures, $\langle W_{\rm diss} \rangle$ thus saturates to a finite constant for RLM and C1CK, but it diverges for critical systems like C2CK and C3CK in a way that is characteristic of the QCP. Here $\Lambda=\Gamma$ for RLM and $T_K$ for CMCK. We set $B=0$.}
    \label{fig:Wdiss_sudden}
\end{figure}

%%%%%%%%%%%%%%%

\noindent\textit{Relation between $\partial_T\langle W_{\rm diss}^3\rangle$ and thermodynamic entropy.-} 
Fig.~\ref{fig:sudden} shows a remarkable similarity in the structure of the CMCK phase diagrams for $\partial_T\langle W_{\rm diss}^3\rangle \equiv\partial_T\kappa^3_W$ in the sudden limit and the impurity contribution to the total thermodynamic entropy, $\Delta S$. This connection can be made precise by considering a local Maxwell relation. Since the entropy in the $(B,T)$ plane gives characteristic information on the renormalization group (RG) fixed points and crossover energy scales, we see that the quantum work statistics (starting at $\kappa^3_W$ where coherence effects first enter) inherit these useful diagnostic properties.

First, let us write $\hat{H}_{\mathcal{SE}}=J \hat{O}_{\mathcal{SE}}$ where $\hat{O}_{\mathcal{SE}}=\hat{S}_d^+\hat{s}^-_c + \hat{S}_d^-\hat{s}^+_c$ for CMCK models. With $F(T,J)$ the $J$- and $T$-dependent free energy, consider a quasistatic process $J:J \to 0$, keeping $(B,T)$ fixed. 
%We are interested in the entropy change for this process -- that is, the difference in total entropy of two equilibrium systems with couplings $J$ and $J=0$. This entropy change is  related to the impurity contribution to the total entropy $\Delta S$ plotted in the lower panel of Fig.~\ref{fig:sudden}. 
The entropy change for this process  -- that is, the difference in total entropy of two equilibrium systems with couplings $J$ and $J=0$ -- is related to $\Delta S$ plotted in the lower panel of Fig.~\ref{fig:sudden}.

The entropy can be obtained from the free energy as $S=-\partial_T F(T,J)$. On the other hand, $\partial_J F(T,J) =\langle\hat{O}_{\mathcal{SE}}\rangle$ such that from Eq.~\ref{eq:3nd_cu} we find $\kappa_W^3=-\tfrac{1}{2} JA^2\partial_J F(T,J)$. Since  $\frac{\partial^2 F}{\partial J \partial T}=\frac{\partial^2 F}{\partial T \partial J }$ we have the Maxwell relation $\partial_J S = - \partial_T \langle \hat{O}_{\mathcal{SE}}\rangle$. Therefore $\partial_T\kappa^3_W$ is related to changes in entropy. Crossovers in $\Delta S$ (e.g.~between fixed points) correspond to minima or maxima in $\partial_T\kappa^3_W$, as observed  in the middle panel of  Fig.~\ref{fig:sudden}. This relation holds in the sudden limit at LR.

%Note that $T_K\equiv T_K(J)\sim J e^{-1/\nu J}$, which in the scaling regime implies that a change in S with J can be related to a change in S with TK. Hence a crossover in S vs B/TK or T/TK corresponds to a peak in $\partial_J S$ and hence a peak in $\partial_T \langle W^3_{\rm diss}\rangle$

%%%%%%%%%%%%%

\noindent\textit{Behavior of $\partial_T\langle W_{\rm diss}^3\rangle$ away from the sudden limit.--}
For the C2CK case, Eq.~\ref{eq:2CK} can be used to find the full field-theory prediction for $\partial_T\langle W_{\rm diss}^3\rangle \equiv\partial_T\kappa^3_W$ for $\tau T \gg 1$,
\be
\frac{1}{A^2}\frac{d\kappa_W^3}{dT} = \frac{T}{ \Lambda}\left[ \frac{2}{(\tau T)^2}\left(\frac{\pi\tau T}{\sinh(\pi\tau T)} - 1 \right)\right].
\label{eq:2ck_dhdt}
\ee
As shown in Fig.~\ref{fig:W3} by the dashed/dotted lines, $\partial_T\kappa^3_W$ displays a peak versus temperature. While the middle panel of Fig.~\ref{fig:sudden} shows this peak in the sudden limit $\tau \to 0$ at $T=T_K$ (horizontal black stripe), Eq.~\ref{eq:2ck_dhdt} indicates that this peak evolves to finite $\tau$ by shifting to $T=\tau^{-1}$ with a height that scales as $\tau^{-1}$. As we approach the adiabatic limit, this energy scale $1/\tau \to 0$, which shows that $\tau$ acts like an inverse energy scale to the QCP. Note that the sudden limit is \textit{not} captured by Eq.~\ref{eq:2ck_dhdt}.

\begin{figure}[t]
\centering
\includegraphics[width=\columnwidth]{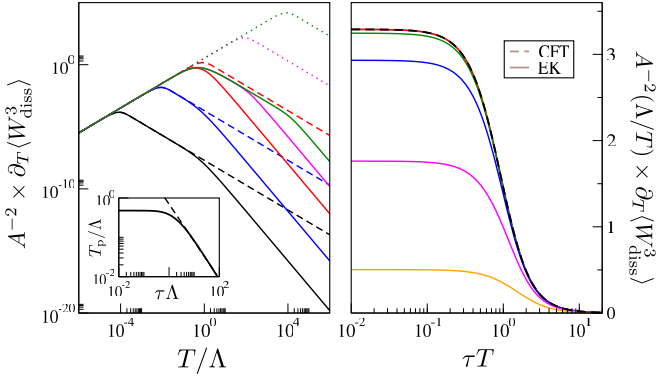}
\caption{Evolution of the third moment of the dissipated work $\langle W^3_{\rm diss}\rangle$ for the C2CK model, plotted as $\partial_T \langle W^3_{\rm diss}\rangle$ vs $T/\Lambda$ (left) and $\Lambda/T\times \partial_T \langle W^3_{\rm diss}\rangle$ vs $\tau T$ (right), where $\Lambda=T_K$ is the cut-off here. The CFT prediction (dashed/dotted lines, Eq.~\ref{eq:2ck_dhdt}) agrees with the full EK solution (solid lines) in the scaling regime $\tau \gg 1/T_K$. Inset shows the temperature $T_p$ of the peak in $\partial_T \langle W^3_{\rm diss}\rangle$, which saturates at $T_p \sim T_K$ for $\tau \ll  1/T_K$.  Plotted for $\tau \Lambda= 10^n$ with $n=4,2,0,-2,-4$ for black, blue, red, magenta and green lines  in the left panel; and $T/\Lambda=10^m$ with $m=-3,-1.5,-1,-0.5,0$ for black, green, blue, magenta and orange lines in the right panel.
}
\label{fig:W3}
\end{figure}

%%%%%%%%%%%%%%%%%%%

\noindent\textit{Emery-Kivelson solution for C2CK.--} 
The low-energy (Fermi liquid) crossover of the C2CK model can be obtained exactly using the bosonization methods of Emery and Kivelson (EK), see~\cite{emery1992mapping,mitchell2016universality}. The effective model is a non-interacting Majorana resonant level model, 
\be
\hat{H}=\epsilon_d d^{\dag}d+iv_F\int_{-\infty}^{\infty}dx\: \psi^{\dagger}\partial_x\psi + \tilde{J}(\psi(0) + \psi^{\dag}(0))(d^{\dag}-d),
\ee
where $d$ is a spinless fermion operator for the impurity and $\psi(x)$ is a field operator for 1d conduction electrons at position $x$ (with $x=0$ the impurity position). Here $v_F$ is the Fermi velocity and $\tilde{J}= J/\sqrt{2\pi a}$ with $a$ the lattice constant. The model is exactly solvable, and in  particular for $\epsilon_d=0$ we have~\cite{emery1992mapping},
\be    
    {\rm{Im}}\chi(\omega) = \frac{1}{2}\tanh{\left(\frac{\omega}{2T}\right)}\frac{\Gamma}{\omega^2+\Gamma^2},
    \label{eq:mrlm_chi}
\ee
with $\Gamma\equiv J^2/(\pi v_Fa)$ the level width. In the original Kondo model, $T_K$ plays the role of $\Gamma$. Moments of 
the dissipated work then follow from Eq.~\ref{eq:cumulants}.

These exact analytic results are plotted as the solid lines in  Fig.~\ref{fig:W3}. Note in particular that the EK solution for C2CK captures the full crossover of the work statistics in $\tau T$, including the sudden limit. The peak in $\partial_T \langle W_{\rm{diss}}^3\rangle$ (left panel, inset) scales as $\tau^{-1}$ for $\tau \Lambda \gg 1$ but correctly saturates for $\tau \Lambda \lesssim 1$, with $\Lambda=\Gamma$ in the EK model. NRG results for the bare C2CK model show the same saturation behavior (except with $\Lambda=T_K$ as expected). The CFT results (Fig.~\ref{fig:W3} dashed lines) agree perfectly with the EK prediction in the scaling regime $T \ll \Lambda$, however the crossover to the sudden limit is not recovered (dotted lines) due to the finite CFT cut-off $\Lambda$.

%%%%%%%%%%%%%%%%%%

\noindent\textit{NRG calculations.--} Our NRG~\cite{wilson1975renormalization,*bulla2008numerical} results for the $M$-channel charge-Kondo models %presented here 
were obtained 
%using Wilson's NRG method \cite{wilson1975renormalization,*bulla2008numerical}, 
utilizing the interleaved Wilson chain technique \cite{mitchell2014generalized,*stadler2016interleaved}, with dynamical quantities obtained using the full density matrix approach \cite{weichselbaum2007sum}. Moments of the dissipated work distribution shown in Fig.~2  involved calculation of the finite-temperature, real-frequency retarded impurity dynamical spin susceptibility, $\chi(\omega,T)=\langle \langle \hat{S}^z_d ; \hat{S}^z_d\rangle\rangle_{\omega,T}$. For Fig.~3 in the sudden quench limit, the static susceptibility $d\langle \hat{S}^z_d\rangle/dB$ and $d\langle \hat{H}_{\mathcal{SE}}\rangle/dT$ were calculated using the differentiable-NRG methodology \cite{rigo2022automatic}. Throughout, we used an NRG discretization parameter $\Lambda=2.5$, combining the results of $N_z=2$ calculations, and kept $N_k=5000$, $12000$ and $40000$ states for the $M=1,2$ and $3$ channel models, respectively. We used $J=0.08D$ in terms of the conduction electron bandwidth $D$, yielding a Kondo temperature $T_K = 10^{-10}D \equiv \Lambda$% which serves as our UV cut-off
. Results presented are therefore all in the %fully 
universal regime. \\

%%%%%%%%%%%

\end{acknowledgments}

%%%%%%%%%%%%%%%%%%%%
%%%%%%%%%%%%%%%%%%%%

\bibliography{bibliography}

\begin{thebibliography}{65}%
\makeatletter
\providecommand \@ifxundefined [1]{%
 \@ifx{#1\undefined}
}%
\providecommand \@ifnum [1]{%
 \ifnum #1\expandafter \@firstoftwo
 \else \expandafter \@secondoftwo
 \fi
}%
\providecommand \@ifx [1]{%
 \ifx #1\expandafter \@firstoftwo
 \else \expandafter \@secondoftwo
 \fi
}%
\providecommand \natexlab [1]{#1}%
\providecommand \enquote  [1]{``#1''}%
\providecommand \bibnamefont  [1]{#1}%
\providecommand \bibfnamefont [1]{#1}%
\providecommand \citenamefont [1]{#1}%
\providecommand \href@noop [0]{\@secondoftwo}%
\providecommand \href [0]{\begingroup \@sanitize@url \@href}%
\providecommand \@href[1]{\@@startlink{#1}\@@href}%
\providecommand \@@href[1]{\endgroup#1\@@endlink}%
\providecommand \@sanitize@url [0]{\catcode `\\12\catcode `\$12\catcode `\&12\catcode `\#12\catcode `\^12\catcode `\_12\catcode `\%12\relax}%
\providecommand \@@startlink[1]{}%
\providecommand \@@endlink[0]{}%
\providecommand \url  [0]{\begingroup\@sanitize@url \@url }%
\providecommand \@url [1]{\endgroup\@href {#1}{\urlprefix }}%
\providecommand \urlprefix  [0]{URL }%
\providecommand \Eprint [0]{\href }%
\providecommand \doibase [0]{https://doi.org/}%
\providecommand \selectlanguage [0]{\@gobble}%
\providecommand \bibinfo  [0]{\@secondoftwo}%
\providecommand \bibfield  [0]{\@secondoftwo}%
\providecommand \translation [1]{[#1]}%
\providecommand \BibitemOpen [0]{}%
\providecommand \bibitemStop [0]{}%
\providecommand \bibitemNoStop [0]{.\EOS\space}%
\providecommand \EOS [0]{\spacefactor3000\relax}%
\providecommand \BibitemShut  [1]{\csname bibitem#1\endcsname}%
\let\auto@bib@innerbib\@empty
%</preamble>
\bibitem [{\citenamefont {Sachdev}(1999)}]{sachdev}%
  \BibitemOpen
  \bibfield  {author} {\bibinfo {author} {\bibfnamefont {S.}~\bibnamefont {Sachdev}},\ }\href@noop {} {\emph {\bibinfo {title} {Quantum Phase Transitions}}}\ (\bibinfo  {publisher} {Cambridge University Press},\ \bibinfo {year} {1999})\BibitemShut {NoStop}%
\bibitem [{\citenamefont {Kibble}(1976)}]{kibble1976topology}%
  \BibitemOpen
  \bibfield  {author} {\bibinfo {author} {\bibfnamefont {T.~W.}\ \bibnamefont {Kibble}},\ }\bibfield  {title} {\bibinfo {title} {Topology of cosmic domains and strings},\ }\href@noop {} {\bibfield  {journal} {\bibinfo  {journal} {Journal of Physics A: Mathematical and General}\ }\textbf {\bibinfo {volume} {9}},\ \bibinfo {pages} {1387} (\bibinfo {year} {1976})}\BibitemShut {NoStop}%
\bibitem [{\citenamefont {Zurek}(1985)}]{zurek1985cosmological}%
  \BibitemOpen
  \bibfield  {author} {\bibinfo {author} {\bibfnamefont {W.~H.}\ \bibnamefont {Zurek}},\ }\bibfield  {title} {\bibinfo {title} {Cosmological experiments in superfluid helium?},\ }\href@noop {} {\bibfield  {journal} {\bibinfo  {journal} {Nature}\ }\textbf {\bibinfo {volume} {317}},\ \bibinfo {pages} {505} (\bibinfo {year} {1985})}\BibitemShut {NoStop}%
\bibitem [{\citenamefont {Calabrese}\ and\ \citenamefont {Cardy}(2016)}]{calabrese2016quantum}%
  \BibitemOpen
  \bibfield  {author} {\bibinfo {author} {\bibfnamefont {P.}~\bibnamefont {Calabrese}}\ and\ \bibinfo {author} {\bibfnamefont {J.}~\bibnamefont {Cardy}},\ }\bibfield  {title} {\bibinfo {title} {Quantum quenches in 1+ 1 dimensional conformal field theories},\ }\href@noop {} {\bibfield  {journal} {\bibinfo  {journal} {Journal of Statistical Mechanics: Theory and Experiment}\ }\textbf {\bibinfo {volume} {2016}},\ \bibinfo {pages} {064003} (\bibinfo {year} {2016})}\BibitemShut {NoStop}%
\bibitem [{\citenamefont {Silva}(2008)}]{silva2008statistics}%
  \BibitemOpen
  \bibfield  {author} {\bibinfo {author} {\bibfnamefont {A.}~\bibnamefont {Silva}},\ }\bibfield  {title} {\bibinfo {title} {Statistics of the work done on a quantum critical system by quenching a control parameter},\ }\href@noop {} {\bibfield  {journal} {\bibinfo  {journal} {Physical review letters}\ }\textbf {\bibinfo {volume} {101}},\ \bibinfo {pages} {120603} (\bibinfo {year} {2008})}\BibitemShut {NoStop}%
\bibitem [{\citenamefont {Fei}\ \emph {et~al.}(2020)\citenamefont {Fei}, \citenamefont {Freitas}, \citenamefont {Cavina}, \citenamefont {Quan},\ and\ \citenamefont {Esposito}}]{fei2020work}%
  \BibitemOpen
  \bibfield  {author} {\bibinfo {author} {\bibfnamefont {Z.}~\bibnamefont {Fei}}, \bibinfo {author} {\bibfnamefont {N.}~\bibnamefont {Freitas}}, \bibinfo {author} {\bibfnamefont {V.}~\bibnamefont {Cavina}}, \bibinfo {author} {\bibfnamefont {H.}~\bibnamefont {Quan}},\ and\ \bibinfo {author} {\bibfnamefont {M.}~\bibnamefont {Esposito}},\ }\bibfield  {title} {\bibinfo {title} {Work statistics across a quantum phase transition},\ }\href@noop {} {\bibfield  {journal} {\bibinfo  {journal} {Physical {R}eview {L}etters}\ }\textbf {\bibinfo {volume} {124}},\ \bibinfo {pages} {170603} (\bibinfo {year} {2020})}\BibitemShut {NoStop}%
\bibitem [{\citenamefont {Rossini}\ and\ \citenamefont {Vicari}(2021)}]{rossini2021coherent}%
  \BibitemOpen
  \bibfield  {author} {\bibinfo {author} {\bibfnamefont {D.}~\bibnamefont {Rossini}}\ and\ \bibinfo {author} {\bibfnamefont {E.}~\bibnamefont {Vicari}},\ }\bibfield  {title} {\bibinfo {title} {Coherent and dissipative dynamics at quantum phase transitions},\ }\href@noop {} {\bibfield  {journal} {\bibinfo  {journal} {Phys. Rep.}\ }\textbf {\bibinfo {volume} {936}},\ \bibinfo {pages} {1} (\bibinfo {year} {2021})}\BibitemShut {NoStop}%
\bibitem [{\citenamefont {Jarzynski}(1997)}]{jarzynski1997nonequilibrium}%
  \BibitemOpen
  \bibfield  {author} {\bibinfo {author} {\bibfnamefont {C.}~\bibnamefont {Jarzynski}},\ }\bibfield  {title} {\bibinfo {title} {Nonequilibrium equality for free energy differences},\ }\href@noop {} {\bibfield  {journal} {\bibinfo  {journal} {Phys. Rev. Lett.}\ }\textbf {\bibinfo {volume} {78}},\ \bibinfo {pages} {2690} (\bibinfo {year} {1997})}\BibitemShut {NoStop}%
\bibitem [{\citenamefont {Dorner}\ \emph {et~al.}(2012)\citenamefont {Dorner}, \citenamefont {Goold}, \citenamefont {Cormick}, \citenamefont {Paternostro},\ and\ \citenamefont {Vedral}}]{PhysRevLett.109.160601}%
  \BibitemOpen
  \bibfield  {author} {\bibinfo {author} {\bibfnamefont {R.}~\bibnamefont {Dorner}}, \bibinfo {author} {\bibfnamefont {J.}~\bibnamefont {Goold}}, \bibinfo {author} {\bibfnamefont {C.}~\bibnamefont {Cormick}}, \bibinfo {author} {\bibfnamefont {M.}~\bibnamefont {Paternostro}},\ and\ \bibinfo {author} {\bibfnamefont {V.}~\bibnamefont {Vedral}},\ }\bibfield  {title} {\bibinfo {title} {Emergent thermodynamics in a quenched quantum many-body system},\ }\href@noop {} {\bibfield  {journal} {\bibinfo  {journal} {Phys. Rev. Lett.}\ }\textbf {\bibinfo {volume} {109}},\ \bibinfo {pages} {160601} (\bibinfo {year} {2012})}\BibitemShut {NoStop}%
\bibitem [{\citenamefont {Fusco}\ \emph {et~al.}(2014)\citenamefont {Fusco}, \citenamefont {Pigeon}, \citenamefont {Apollaro}, \citenamefont {Xuereb}, \citenamefont {Mazzola}, \citenamefont {Campisi}, \citenamefont {Ferraro}, \citenamefont {Paternostro},\ and\ \citenamefont {De~Chiara}}]{PhysRevX.4.031029}%
  \BibitemOpen
  \bibfield  {author} {\bibinfo {author} {\bibfnamefont {L.}~\bibnamefont {Fusco}}, \bibinfo {author} {\bibfnamefont {S.}~\bibnamefont {Pigeon}}, \bibinfo {author} {\bibfnamefont {T.~J.~G.}\ \bibnamefont {Apollaro}}, \bibinfo {author} {\bibfnamefont {A.}~\bibnamefont {Xuereb}}, \bibinfo {author} {\bibfnamefont {L.}~\bibnamefont {Mazzola}}, \bibinfo {author} {\bibfnamefont {M.}~\bibnamefont {Campisi}}, \bibinfo {author} {\bibfnamefont {A.}~\bibnamefont {Ferraro}}, \bibinfo {author} {\bibfnamefont {M.}~\bibnamefont {Paternostro}},\ and\ \bibinfo {author} {\bibfnamefont {G.}~\bibnamefont {De~Chiara}},\ }\bibfield  {title} {\bibinfo {title} {Assessing the nonequilibrium thermodynamics in a quenched quantum many-body system via single projective measurements},\ }\href@noop {} {\bibfield  {journal} {\bibinfo  {journal} {Phys. Rev. X}\ }\textbf {\bibinfo {volume} {4}},\ \bibinfo {pages} {031029} (\bibinfo {year} {2014})}\BibitemShut {NoStop}%
\bibitem [{\citenamefont {Goold}\ \emph {et~al.}(2018)\citenamefont {Goold}, \citenamefont {Plastina}, \citenamefont {Gambassi},\ and\ \citenamefont {Silva}}]{goold2018role}%
  \BibitemOpen
  \bibfield  {author} {\bibinfo {author} {\bibfnamefont {J.}~\bibnamefont {Goold}}, \bibinfo {author} {\bibfnamefont {F.}~\bibnamefont {Plastina}}, \bibinfo {author} {\bibfnamefont {A.}~\bibnamefont {Gambassi}},\ and\ \bibinfo {author} {\bibfnamefont {A.}~\bibnamefont {Silva}},\ }\bibfield  {title} {\bibinfo {title} {The role of quantum work statistics in many-body physics},\ }\href@noop {} {\bibfield  {journal} {\bibinfo  {journal} {Thermodynamics in the Quantum Regime: Fundamental Aspects and New Directions}\ ,\ \bibinfo {pages} {317}} (\bibinfo {year} {2018})}\BibitemShut {NoStop}%
\bibitem [{\citenamefont {Mascarenhas}\ \emph {et~al.}(2014)\citenamefont {Mascarenhas}, \citenamefont {Bragan\ifmmode~\mbox{\c{c}}\else \c{c}\fi{}a}, \citenamefont {Dorner}, \citenamefont {Fran\ifmmode \mbox{\c{c}}\else~\c{c}\fi{}a Santos}, \citenamefont {Vedral}, \citenamefont {Modi},\ and\ \citenamefont {Goold}}]{PhysRevE.89.062103}%
  \BibitemOpen
  \bibfield  {author} {\bibinfo {author} {\bibfnamefont {E.}~\bibnamefont {Mascarenhas}}, \bibinfo {author} {\bibfnamefont {H.}~\bibnamefont {Bragan\ifmmode~\mbox{\c{c}}\else \c{c}\fi{}a}}, \bibinfo {author} {\bibfnamefont {R.}~\bibnamefont {Dorner}}, \bibinfo {author} {\bibfnamefont {M.}~\bibnamefont {Fran\ifmmode \mbox{\c{c}}\else~\c{c}\fi{}a Santos}}, \bibinfo {author} {\bibfnamefont {V.}~\bibnamefont {Vedral}}, \bibinfo {author} {\bibfnamefont {K.}~\bibnamefont {Modi}},\ and\ \bibinfo {author} {\bibfnamefont {J.}~\bibnamefont {Goold}},\ }\bibfield  {title} {\bibinfo {title} {Work and quantum phase transitions: Quantum latency},\ }\href {https://doi.org/10.1103/PhysRevE.89.062103} {\bibfield  {journal} {\bibinfo  {journal} {Phys. Rev. E}\ }\textbf {\bibinfo {volume} {89}},\ \bibinfo {pages} {062103} (\bibinfo {year} {2014})}\BibitemShut {NoStop}%
\bibitem [{\citenamefont {Suomela}\ \emph {et~al.}(2014)\citenamefont {Suomela}, \citenamefont {Solinas}, \citenamefont {Pekola}, \citenamefont {Ankerhold},\ and\ \citenamefont {Ala-Nissila}}]{PhysRevB.90.094304}%
  \BibitemOpen
  \bibfield  {author} {\bibinfo {author} {\bibfnamefont {S.}~\bibnamefont {Suomela}}, \bibinfo {author} {\bibfnamefont {P.}~\bibnamefont {Solinas}}, \bibinfo {author} {\bibfnamefont {J.~P.}\ \bibnamefont {Pekola}}, \bibinfo {author} {\bibfnamefont {J.}~\bibnamefont {Ankerhold}},\ and\ \bibinfo {author} {\bibfnamefont {T.}~\bibnamefont {Ala-Nissila}},\ }\bibfield  {title} {\bibinfo {title} {Moments of work in the two-point measurement protocol for a driven open quantum system},\ }\href {https://doi.org/10.1103/PhysRevB.90.094304} {\bibfield  {journal} {\bibinfo  {journal} {Phys. Rev. B}\ }\textbf {\bibinfo {volume} {90}},\ \bibinfo {pages} {094304} (\bibinfo {year} {2014})}\BibitemShut {NoStop}%
\bibitem [{\citenamefont {Deffner}(2017)}]{deffner2017kibble}%
  \BibitemOpen
  \bibfield  {author} {\bibinfo {author} {\bibfnamefont {S.}~\bibnamefont {Deffner}},\ }\bibfield  {title} {\bibinfo {title} {Kibble-zurek scaling of the irreversible entropy production},\ }\href@noop {} {\bibfield  {journal} {\bibinfo  {journal} {Physical Review E}\ }\textbf {\bibinfo {volume} {96}},\ \bibinfo {pages} {052125} (\bibinfo {year} {2017})}\BibitemShut {NoStop}%
\bibitem [{\citenamefont {Scandi}\ \emph {et~al.}(2020)\citenamefont {Scandi}, \citenamefont {Miller}, \citenamefont {Anders},\ and\ \citenamefont {Perarnau-Llobet}}]{PhysRevResearch.2.023377}%
  \BibitemOpen
  \bibfield  {author} {\bibinfo {author} {\bibfnamefont {M.}~\bibnamefont {Scandi}}, \bibinfo {author} {\bibfnamefont {H.~J.~D.}\ \bibnamefont {Miller}}, \bibinfo {author} {\bibfnamefont {J.}~\bibnamefont {Anders}},\ and\ \bibinfo {author} {\bibfnamefont {M.}~\bibnamefont {Perarnau-Llobet}},\ }\bibfield  {title} {\bibinfo {title} {Quantum work statistics close to equilibrium},\ }\href@noop {} {\bibfield  {journal} {\bibinfo  {journal} {Phys. Rev. Res.}\ }\textbf {\bibinfo {volume} {2}},\ \bibinfo {pages} {023377} (\bibinfo {year} {2020})}\BibitemShut {NoStop}%
\bibitem [{\citenamefont {Van~Vu}\ and\ \citenamefont {Saito}(2022)}]{PhysRevLett.128.010602}%
  \BibitemOpen
  \bibfield  {author} {\bibinfo {author} {\bibfnamefont {T.}~\bibnamefont {Van~Vu}}\ and\ \bibinfo {author} {\bibfnamefont {K.}~\bibnamefont {Saito}},\ }\bibfield  {title} {\bibinfo {title} {Finite-time quantum landauer principle and quantum coherence},\ }\href@noop {} {\bibfield  {journal} {\bibinfo  {journal} {Phys. Rev. Lett.}\ }\textbf {\bibinfo {volume} {128}},\ \bibinfo {pages} {010602} (\bibinfo {year} {2022})}\BibitemShut {NoStop}%
\bibitem [{\citenamefont {Miller}\ \emph {et~al.}(2020)\citenamefont {Miller}, \citenamefont {Guarnieri}, \citenamefont {Mitchison},\ and\ \citenamefont {Goold}}]{PhysRevLett.125.160602}%
  \BibitemOpen
  \bibfield  {author} {\bibinfo {author} {\bibfnamefont {H.~J.~D.}\ \bibnamefont {Miller}}, \bibinfo {author} {\bibfnamefont {G.}~\bibnamefont {Guarnieri}}, \bibinfo {author} {\bibfnamefont {M.~T.}\ \bibnamefont {Mitchison}},\ and\ \bibinfo {author} {\bibfnamefont {J.}~\bibnamefont {Goold}},\ }\bibfield  {title} {\bibinfo {title} {Quantum fluctuations hinder finite-time information erasure near the landauer limit},\ }\href@noop {} {\bibfield  {journal} {\bibinfo  {journal} {Phys. Rev. Lett.}\ }\textbf {\bibinfo {volume} {125}},\ \bibinfo {pages} {160602} (\bibinfo {year} {2020})}\BibitemShut {NoStop}%
\bibitem [{\citenamefont {Zawadzki}\ \emph {et~al.}(2020)\citenamefont {Zawadzki}, \citenamefont {Serra},\ and\ \citenamefont {D'Amico}}]{zawadzki2020work}%
  \BibitemOpen
  \bibfield  {author} {\bibinfo {author} {\bibfnamefont {K.}~\bibnamefont {Zawadzki}}, \bibinfo {author} {\bibfnamefont {R.~M.}\ \bibnamefont {Serra}},\ and\ \bibinfo {author} {\bibfnamefont {I.}~\bibnamefont {D'Amico}},\ }\bibfield  {title} {\bibinfo {title} {Work-distribution quantumness and irreversibility when crossing a quantum phase transition in finite time},\ }\href@noop {} {\bibfield  {journal} {\bibinfo  {journal} {Physical Review Research}\ }\textbf {\bibinfo {volume} {2}},\ \bibinfo {pages} {033167} (\bibinfo {year} {2020})}\BibitemShut {NoStop}%
\bibitem [{\citenamefont {Zawadzki}\ \emph {et~al.}(2023)\citenamefont {Zawadzki}, \citenamefont {Kiely}, \citenamefont {Landi},\ and\ \citenamefont {Campbell}}]{zawadzki2023non}%
  \BibitemOpen
  \bibfield  {author} {\bibinfo {author} {\bibfnamefont {K.}~\bibnamefont {Zawadzki}}, \bibinfo {author} {\bibfnamefont {A.}~\bibnamefont {Kiely}}, \bibinfo {author} {\bibfnamefont {G.~T.}\ \bibnamefont {Landi}},\ and\ \bibinfo {author} {\bibfnamefont {S.}~\bibnamefont {Campbell}},\ }\bibfield  {title} {\bibinfo {title} {Non-gaussian work statistics at finite-time driving},\ }\href@noop {} {\bibfield  {journal} {\bibinfo  {journal} {Physical {R}eview {A}}\ }\textbf {\bibinfo {volume} {107}},\ \bibinfo {pages} {012209} (\bibinfo {year} {2023})}\BibitemShut {NoStop}%
\bibitem [{\citenamefont {Saira}\ \emph {et~al.}(2012)\citenamefont {Saira}, \citenamefont {Yoon}, \citenamefont {Tanttu}, \citenamefont {M{\"o}tt{\"o}nen}, \citenamefont {Averin},\ and\ \citenamefont {Pekola}}]{saira2012test}%
  \BibitemOpen
  \bibfield  {author} {\bibinfo {author} {\bibfnamefont {O.-P.}\ \bibnamefont {Saira}}, \bibinfo {author} {\bibfnamefont {Y.}~\bibnamefont {Yoon}}, \bibinfo {author} {\bibfnamefont {T.}~\bibnamefont {Tanttu}}, \bibinfo {author} {\bibfnamefont {M.}~\bibnamefont {M{\"o}tt{\"o}nen}}, \bibinfo {author} {\bibfnamefont {D.}~\bibnamefont {Averin}},\ and\ \bibinfo {author} {\bibfnamefont {J.~P.}\ \bibnamefont {Pekola}},\ }\bibfield  {title} {\bibinfo {title} {Test of the jarzynski and crooks fluctuation relations in an electronic system},\ }\href@noop {} {\bibfield  {journal} {\bibinfo  {journal} {Phys. Rev. Lett.}\ }\textbf {\bibinfo {volume} {109}},\ \bibinfo {pages} {180601} (\bibinfo {year} {2012})}\BibitemShut {NoStop}%
\bibitem [{\citenamefont {Koski}\ \emph {et~al.}(2013)\citenamefont {Koski}, \citenamefont {Sagawa}, \citenamefont {Saira}, \citenamefont {Yoon}, \citenamefont {Kutvonen}, \citenamefont {Solinas}, \citenamefont {M{\"o}tt{\"o}nen}, \citenamefont {Ala-Nissila},\ and\ \citenamefont {Pekola}}]{koski2013distribution}%
  \BibitemOpen
  \bibfield  {author} {\bibinfo {author} {\bibfnamefont {J.}~\bibnamefont {Koski}}, \bibinfo {author} {\bibfnamefont {T.}~\bibnamefont {Sagawa}}, \bibinfo {author} {\bibfnamefont {O.}~\bibnamefont {Saira}}, \bibinfo {author} {\bibfnamefont {Y.}~\bibnamefont {Yoon}}, \bibinfo {author} {\bibfnamefont {A.}~\bibnamefont {Kutvonen}}, \bibinfo {author} {\bibfnamefont {P.}~\bibnamefont {Solinas}}, \bibinfo {author} {\bibfnamefont {M.}~\bibnamefont {M{\"o}tt{\"o}nen}}, \bibinfo {author} {\bibfnamefont {T.}~\bibnamefont {Ala-Nissila}},\ and\ \bibinfo {author} {\bibfnamefont {J.}~\bibnamefont {Pekola}},\ }\bibfield  {title} {\bibinfo {title} {Distribution of entropy production in a single-electron box},\ }\href@noop {} {\bibfield  {journal} {\bibinfo  {journal} {Nat. Phys.}\ }\textbf {\bibinfo {volume} {9}},\ \bibinfo {pages} {644} (\bibinfo {year} {2013})}\BibitemShut {NoStop}%
\bibitem [{\citenamefont {Hofmann}\ \emph {et~al.}(2017)\citenamefont {Hofmann}, \citenamefont {Maisi}, \citenamefont {Basset}, \citenamefont {Reichl}, \citenamefont {Wegscheider}, \citenamefont {Ihn}, \citenamefont {Ensslin},\ and\ \citenamefont {Jarzynski}}]{hofmann2017heat}%
  \BibitemOpen
  \bibfield  {author} {\bibinfo {author} {\bibfnamefont {A.}~\bibnamefont {Hofmann}}, \bibinfo {author} {\bibfnamefont {V.~F.}\ \bibnamefont {Maisi}}, \bibinfo {author} {\bibfnamefont {J.}~\bibnamefont {Basset}}, \bibinfo {author} {\bibfnamefont {C.}~\bibnamefont {Reichl}}, \bibinfo {author} {\bibfnamefont {W.}~\bibnamefont {Wegscheider}}, \bibinfo {author} {\bibfnamefont {T.}~\bibnamefont {Ihn}}, \bibinfo {author} {\bibfnamefont {K.}~\bibnamefont {Ensslin}},\ and\ \bibinfo {author} {\bibfnamefont {C.}~\bibnamefont {Jarzynski}},\ }\bibfield  {title} {\bibinfo {title} {Heat dissipation and fluctuations in a driven quantum dot},\ }\href@noop {} {\bibfield  {journal} {\bibinfo  {journal} {Phys. Stat. Sol. B}\ }\textbf {\bibinfo {volume} {254}},\ \bibinfo {pages} {1600546} (\bibinfo {year} {2017})}\BibitemShut {NoStop}%
\bibitem [{\citenamefont {Hofmann}\ \emph {et~al.}(2016)\citenamefont {Hofmann}, \citenamefont {Maisi}, \citenamefont {R{\"o}ssler}, \citenamefont {Basset}, \citenamefont {Kr{\"a}henmann}, \citenamefont {M{\"a}rki}, \citenamefont {Ihn}, \citenamefont {Ensslin}, \citenamefont {Reichl},\ and\ \citenamefont {Wegscheider}}]{hofmann2016equilibrium}%
  \BibitemOpen
  \bibfield  {author} {\bibinfo {author} {\bibfnamefont {A.}~\bibnamefont {Hofmann}}, \bibinfo {author} {\bibfnamefont {V.~F.}\ \bibnamefont {Maisi}}, \bibinfo {author} {\bibfnamefont {C.}~\bibnamefont {R{\"o}ssler}}, \bibinfo {author} {\bibfnamefont {J.}~\bibnamefont {Basset}}, \bibinfo {author} {\bibfnamefont {T.}~\bibnamefont {Kr{\"a}henmann}}, \bibinfo {author} {\bibfnamefont {P.}~\bibnamefont {M{\"a}rki}}, \bibinfo {author} {\bibfnamefont {T.}~\bibnamefont {Ihn}}, \bibinfo {author} {\bibfnamefont {K.}~\bibnamefont {Ensslin}}, \bibinfo {author} {\bibfnamefont {C.}~\bibnamefont {Reichl}},\ and\ \bibinfo {author} {\bibfnamefont {W.}~\bibnamefont {Wegscheider}},\ }\bibfield  {title} {\bibinfo {title} {Equilibrium free energy measurement of a confined electron driven out of equilibrium},\ }\href@noop {} {\bibfield  {journal} {\bibinfo  {journal} {Phys. Rev. B}\ }\textbf {\bibinfo {volume} {93}},\ \bibinfo {pages} {035425} (\bibinfo {year} {2016})}\BibitemShut {NoStop}%
\bibitem [{\citenamefont {Barker}\ \emph {et~al.}(2022)\citenamefont {Barker}, \citenamefont {Scandi}, \citenamefont {Lehmann}, \citenamefont {Thelander}, \citenamefont {Dick}, \citenamefont {Perarnau-Llobet},\ and\ \citenamefont {Maisi}}]{barker2022experimental}%
  \BibitemOpen
  \bibfield  {author} {\bibinfo {author} {\bibfnamefont {D.}~\bibnamefont {Barker}}, \bibinfo {author} {\bibfnamefont {M.}~\bibnamefont {Scandi}}, \bibinfo {author} {\bibfnamefont {S.}~\bibnamefont {Lehmann}}, \bibinfo {author} {\bibfnamefont {C.}~\bibnamefont {Thelander}}, \bibinfo {author} {\bibfnamefont {K.~A.}\ \bibnamefont {Dick}}, \bibinfo {author} {\bibfnamefont {M.}~\bibnamefont {Perarnau-Llobet}},\ and\ \bibinfo {author} {\bibfnamefont {V.~F.}\ \bibnamefont {Maisi}},\ }\bibfield  {title} {\bibinfo {title} {Experimental verification of the work fluctuation-dissipation relation for information-to-work conversion},\ }\href@noop {} {\bibfield  {journal} {\bibinfo  {journal} {Phys. Rev. Lett.}\ }\textbf {\bibinfo {volume} {128}},\ \bibinfo {pages} {040602} (\bibinfo {year} {2022})}\BibitemShut {NoStop}%
\bibitem [{\citenamefont {Moreira}\ \emph {et~al.}(2023)\citenamefont {Moreira}, \citenamefont {Samuelsson},\ and\ \citenamefont {Potts}}]{PhysRevLett.131.220405}%
  \BibitemOpen
  \bibfield  {author} {\bibinfo {author} {\bibfnamefont {S.~V.}\ \bibnamefont {Moreira}}, \bibinfo {author} {\bibfnamefont {P.}~\bibnamefont {Samuelsson}},\ and\ \bibinfo {author} {\bibfnamefont {P.~P.}\ \bibnamefont {Potts}},\ }\bibfield  {title} {\bibinfo {title} {Stochastic thermodynamics of a quantum dot coupled to a finite-size reservoir},\ }\href@noop {} {\bibfield  {journal} {\bibinfo  {journal} {Phys. Rev. Lett.}\ }\textbf {\bibinfo {volume} {131}},\ \bibinfo {pages} {220405} (\bibinfo {year} {2023})}\BibitemShut {NoStop}%
\bibitem [{\citenamefont {Potok}\ \emph {et~al.}(2007)\citenamefont {Potok}, \citenamefont {Rau}, \citenamefont {Shtrikman}, \citenamefont {Oreg},\ and\ \citenamefont {Goldhaber-Gordon}}]{potok2007observation}%
  \BibitemOpen
  \bibfield  {author} {\bibinfo {author} {\bibfnamefont {R.}~\bibnamefont {Potok}}, \bibinfo {author} {\bibfnamefont {I.}~\bibnamefont {Rau}}, \bibinfo {author} {\bibfnamefont {H.}~\bibnamefont {Shtrikman}}, \bibinfo {author} {\bibfnamefont {Y.}~\bibnamefont {Oreg}},\ and\ \bibinfo {author} {\bibfnamefont {D.}~\bibnamefont {Goldhaber-Gordon}},\ }\bibfield  {title} {\bibinfo {title} {Observation of the two-channel kondo effect},\ }\href@noop {} {\bibfield  {journal} {\bibinfo  {journal} {Nature}\ }\textbf {\bibinfo {volume} {446}},\ \bibinfo {pages} {167} (\bibinfo {year} {2007})}\BibitemShut {NoStop}%
\bibitem [{\citenamefont {Keller}\ \emph {et~al.}(2015)\citenamefont {Keller}, \citenamefont {Peeters}, \citenamefont {Moca}, \citenamefont {Weymann}, \citenamefont {Mahalu}, \citenamefont {Umansky}, \citenamefont {Zar{\'a}nd},\ and\ \citenamefont {Goldhaber-Gordon}}]{keller2015universal}%
  \BibitemOpen
  \bibfield  {author} {\bibinfo {author} {\bibfnamefont {A.}~\bibnamefont {Keller}}, \bibinfo {author} {\bibfnamefont {L.}~\bibnamefont {Peeters}}, \bibinfo {author} {\bibfnamefont {C.}~\bibnamefont {Moca}}, \bibinfo {author} {\bibfnamefont {I.}~\bibnamefont {Weymann}}, \bibinfo {author} {\bibfnamefont {D.}~\bibnamefont {Mahalu}}, \bibinfo {author} {\bibfnamefont {V.}~\bibnamefont {Umansky}}, \bibinfo {author} {\bibfnamefont {G.}~\bibnamefont {Zar{\'a}nd}},\ and\ \bibinfo {author} {\bibfnamefont {D.}~\bibnamefont {Goldhaber-Gordon}},\ }\bibfield  {title} {\bibinfo {title} {Universal fermi liquid crossover and quantum criticality in a mesoscopic system},\ }\href@noop {} {\bibfield  {journal} {\bibinfo  {journal} {Nature}\ }\textbf {\bibinfo {volume} {526}},\ \bibinfo {pages} {237} (\bibinfo {year} {2015})}\BibitemShut {NoStop}%
\bibitem [{\citenamefont {Mebrahtu}\ \emph {et~al.}(2012)\citenamefont {Mebrahtu}, \citenamefont {Borzenets}, \citenamefont {Liu}, \citenamefont {Zheng}, \citenamefont {Bomze}, \citenamefont {Smirnov}, \citenamefont {Baranger},\ and\ \citenamefont {Finkelstein}}]{Mebrahtu_2012}%
  \BibitemOpen
  \bibfield  {author} {\bibinfo {author} {\bibfnamefont {H.~T.}\ \bibnamefont {Mebrahtu}}, \bibinfo {author} {\bibfnamefont {I.~V.}\ \bibnamefont {Borzenets}}, \bibinfo {author} {\bibfnamefont {D.~E.}\ \bibnamefont {Liu}}, \bibinfo {author} {\bibfnamefont {H.}~\bibnamefont {Zheng}}, \bibinfo {author} {\bibfnamefont {Y.~V.}\ \bibnamefont {Bomze}}, \bibinfo {author} {\bibfnamefont {A.~I.}\ \bibnamefont {Smirnov}}, \bibinfo {author} {\bibfnamefont {H.~U.}\ \bibnamefont {Baranger}},\ and\ \bibinfo {author} {\bibfnamefont {G.}~\bibnamefont {Finkelstein}},\ }\bibfield  {title} {\bibinfo {title} {Quantum phase transition in a resonant level coupled to interacting leads},\ }\href {https://doi.org/10.1038/nature11265} {\bibfield  {journal} {\bibinfo  {journal} {Nature}\ }\textbf {\bibinfo {volume} {488}},\ \bibinfo {pages} {61–64} (\bibinfo {year} {2012})}\BibitemShut {NoStop}%
\bibitem [{\citenamefont {Mebrahtu}\ \emph {et~al.}(2013)\citenamefont {Mebrahtu}, \citenamefont {Borzenets}, \citenamefont {Zheng}, \citenamefont {Bomze}, \citenamefont {Smirnov}, \citenamefont {Florens}, \citenamefont {Baranger},\ and\ \citenamefont {Finkelstein}}]{Mebrahtu_2013}%
  \BibitemOpen
  \bibfield  {author} {\bibinfo {author} {\bibfnamefont {H.~T.}\ \bibnamefont {Mebrahtu}}, \bibinfo {author} {\bibfnamefont {I.~V.}\ \bibnamefont {Borzenets}}, \bibinfo {author} {\bibfnamefont {H.}~\bibnamefont {Zheng}}, \bibinfo {author} {\bibfnamefont {Y.~V.}\ \bibnamefont {Bomze}}, \bibinfo {author} {\bibfnamefont {A.~I.}\ \bibnamefont {Smirnov}}, \bibinfo {author} {\bibfnamefont {S.}~\bibnamefont {Florens}}, \bibinfo {author} {\bibfnamefont {H.~U.}\ \bibnamefont {Baranger}},\ and\ \bibinfo {author} {\bibfnamefont {G.}~\bibnamefont {Finkelstein}},\ }\bibfield  {title} {\bibinfo {title} {Observation of majorana quantum critical behaviour in a resonant level coupled to a dissipative environment},\ }\href {https://doi.org/10.1038/nphys2735} {\bibfield  {journal} {\bibinfo  {journal} {Nat. Phys.}\ }\textbf {\bibinfo {volume} {9}},\ \bibinfo {pages} {732–737} (\bibinfo {year} {2013})}\BibitemShut {NoStop}%
\bibitem [{\citenamefont {Iftikhar}\ \emph {et~al.}(2015)\citenamefont {Iftikhar}, \citenamefont {Jezouin}, \citenamefont {Anthore}, \citenamefont {Gennser}, \citenamefont {Parmentier}, \citenamefont {Cavanna},\ and\ \citenamefont {Pierre}}]{iftikhar2015two}%
  \BibitemOpen
  \bibfield  {author} {\bibinfo {author} {\bibfnamefont {Z.}~\bibnamefont {Iftikhar}}, \bibinfo {author} {\bibfnamefont {S.}~\bibnamefont {Jezouin}}, \bibinfo {author} {\bibfnamefont {A.}~\bibnamefont {Anthore}}, \bibinfo {author} {\bibfnamefont {U.}~\bibnamefont {Gennser}}, \bibinfo {author} {\bibfnamefont {F.}~\bibnamefont {Parmentier}}, \bibinfo {author} {\bibfnamefont {A.}~\bibnamefont {Cavanna}},\ and\ \bibinfo {author} {\bibfnamefont {F.}~\bibnamefont {Pierre}},\ }\bibfield  {title} {\bibinfo {title} {Two-channel kondo effect and renormalization flow with macroscopic quantum charge states},\ }\href@noop {} {\bibfield  {journal} {\bibinfo  {journal} {Nature}\ }\textbf {\bibinfo {volume} {526}},\ \bibinfo {pages} {233} (\bibinfo {year} {2015})}\BibitemShut {NoStop}%
\bibitem [{\citenamefont {Mitchell}\ \emph {et~al.}(2016)\citenamefont {Mitchell}, \citenamefont {Landau}, \citenamefont {Fritz},\ and\ \citenamefont {Sela}}]{mitchell2016universality}%
  \BibitemOpen
  \bibfield  {author} {\bibinfo {author} {\bibfnamefont {A.~K.}\ \bibnamefont {Mitchell}}, \bibinfo {author} {\bibfnamefont {L.}~\bibnamefont {Landau}}, \bibinfo {author} {\bibfnamefont {L.}~\bibnamefont {Fritz}},\ and\ \bibinfo {author} {\bibfnamefont {E.}~\bibnamefont {Sela}},\ }\bibfield  {title} {\bibinfo {title} {Universality and scaling in a charge two-channel kondo device},\ }\href@noop {} {\bibfield  {journal} {\bibinfo  {journal} {Physical review letters}\ }\textbf {\bibinfo {volume} {116}},\ \bibinfo {pages} {157202} (\bibinfo {year} {2016})}\BibitemShut {NoStop}%
\bibitem [{\citenamefont {Iftikhar}\ \emph {et~al.}(2018)\citenamefont {Iftikhar}, \citenamefont {Anthore}, \citenamefont {Mitchell}, \citenamefont {Parmentier}, \citenamefont {Gennser}, \citenamefont {Ouerghi}, \citenamefont {Cavanna}, \citenamefont {Mora}, \citenamefont {Simon},\ and\ \citenamefont {Pierre}}]{iftikhar2018tunable}%
  \BibitemOpen
  \bibfield  {author} {\bibinfo {author} {\bibfnamefont {Z.}~\bibnamefont {Iftikhar}}, \bibinfo {author} {\bibfnamefont {A.}~\bibnamefont {Anthore}}, \bibinfo {author} {\bibfnamefont {A.}~\bibnamefont {Mitchell}}, \bibinfo {author} {\bibfnamefont {F.}~\bibnamefont {Parmentier}}, \bibinfo {author} {\bibfnamefont {U.}~\bibnamefont {Gennser}}, \bibinfo {author} {\bibfnamefont {A.}~\bibnamefont {Ouerghi}}, \bibinfo {author} {\bibfnamefont {A.}~\bibnamefont {Cavanna}}, \bibinfo {author} {\bibfnamefont {C.}~\bibnamefont {Mora}}, \bibinfo {author} {\bibfnamefont {P.}~\bibnamefont {Simon}},\ and\ \bibinfo {author} {\bibfnamefont {F.}~\bibnamefont {Pierre}},\ }\bibfield  {title} {\bibinfo {title} {Tunable quantum criticality and super-ballistic transport in a “charge” kondo circuit},\ }\href@noop {} {\bibfield  {journal} {\bibinfo  {journal} {Science}\ }\textbf {\bibinfo {volume} {360}},\ \bibinfo {pages} {1315} (\bibinfo {year} {2018})}\BibitemShut {NoStop}%
\bibitem [{\citenamefont {Han}\ \emph {et~al.}(2022)\citenamefont {Han}, \citenamefont {Iftikhar}, \citenamefont {Kleeorin}, \citenamefont {Anthore}, \citenamefont {Pierre}, \citenamefont {Meir}, \citenamefont {Mitchell},\ and\ \citenamefont {Sela}}]{han2022fractional}%
  \BibitemOpen
  \bibfield  {author} {\bibinfo {author} {\bibfnamefont {C.}~\bibnamefont {Han}}, \bibinfo {author} {\bibfnamefont {Z.}~\bibnamefont {Iftikhar}}, \bibinfo {author} {\bibfnamefont {Y.}~\bibnamefont {Kleeorin}}, \bibinfo {author} {\bibfnamefont {A.}~\bibnamefont {Anthore}}, \bibinfo {author} {\bibfnamefont {F.}~\bibnamefont {Pierre}}, \bibinfo {author} {\bibfnamefont {Y.}~\bibnamefont {Meir}}, \bibinfo {author} {\bibfnamefont {A.~K.}\ \bibnamefont {Mitchell}},\ and\ \bibinfo {author} {\bibfnamefont {E.}~\bibnamefont {Sela}},\ }\bibfield  {title} {\bibinfo {title} {Fractional entropy of multichannel kondo systems from conductance-charge relations},\ }\href@noop {} {\bibfield  {journal} {\bibinfo  {journal} {Phys. Rev. Lett.}\ }\textbf {\bibinfo {volume} {128}},\ \bibinfo {pages} {146803} (\bibinfo {year} {2022})}\BibitemShut {NoStop}%
\bibitem [{\citenamefont {Pouse}\ \emph {et~al.}(2023)\citenamefont {Pouse}, \citenamefont {Peeters}, \citenamefont {Hsueh}, \citenamefont {Gennser}, \citenamefont {Cavanna}, \citenamefont {Kastner}, \citenamefont {Mitchell},\ and\ \citenamefont {Goldhaber-Gordon}}]{pouse2023quantum}%
  \BibitemOpen
  \bibfield  {author} {\bibinfo {author} {\bibfnamefont {W.}~\bibnamefont {Pouse}}, \bibinfo {author} {\bibfnamefont {L.}~\bibnamefont {Peeters}}, \bibinfo {author} {\bibfnamefont {C.~L.}\ \bibnamefont {Hsueh}}, \bibinfo {author} {\bibfnamefont {U.}~\bibnamefont {Gennser}}, \bibinfo {author} {\bibfnamefont {A.}~\bibnamefont {Cavanna}}, \bibinfo {author} {\bibfnamefont {M.~A.}\ \bibnamefont {Kastner}}, \bibinfo {author} {\bibfnamefont {A.~K.}\ \bibnamefont {Mitchell}},\ and\ \bibinfo {author} {\bibfnamefont {D.}~\bibnamefont {Goldhaber-Gordon}},\ }\bibfield  {title} {\bibinfo {title} {Quantum simulation of an exotic quantum critical point in a two-site charge kondo circuit},\ }\href@noop {} {\bibfield  {journal} {\bibinfo  {journal} {Nature Physics}\ }\textbf {\bibinfo {volume} {19}},\ \bibinfo {pages} {492} (\bibinfo {year} {2023})}\BibitemShut {NoStop}%
\bibitem [{\citenamefont {Karki}\ \emph {et~al.}(2023)\citenamefont {Karki}, \citenamefont {Boulat}, \citenamefont {Pouse}, \citenamefont {Goldhaber-Gordon}, \citenamefont {Mitchell},\ and\ \citenamefont {Mora}}]{karki2023z}%
  \BibitemOpen
  \bibfield  {author} {\bibinfo {author} {\bibfnamefont {D.}~\bibnamefont {Karki}}, \bibinfo {author} {\bibfnamefont {E.}~\bibnamefont {Boulat}}, \bibinfo {author} {\bibfnamefont {W.}~\bibnamefont {Pouse}}, \bibinfo {author} {\bibfnamefont {D.}~\bibnamefont {Goldhaber-Gordon}}, \bibinfo {author} {\bibfnamefont {A.~K.}\ \bibnamefont {Mitchell}},\ and\ \bibinfo {author} {\bibfnamefont {C.}~\bibnamefont {Mora}},\ }\bibfield  {title} {\bibinfo {title} {Z 3 parafermion in the double charge kondo model},\ }\href@noop {} {\bibfield  {journal} {\bibinfo  {journal} {Physical Review Letters}\ }\textbf {\bibinfo {volume} {130}},\ \bibinfo {pages} {146201} (\bibinfo {year} {2023})}\BibitemShut {NoStop}%
\bibitem [{\citenamefont {Piquard}\ \emph {et~al.}(2023)\citenamefont {Piquard}, \citenamefont {Glidic}, \citenamefont {Han}, \citenamefont {Aassime}, \citenamefont {Cavanna}, \citenamefont {Gennser}, \citenamefont {Meir}, \citenamefont {Sela}, \citenamefont {Anthore},\ and\ \citenamefont {Pierre}}]{piquard2023observing}%
  \BibitemOpen
  \bibfield  {author} {\bibinfo {author} {\bibfnamefont {C.}~\bibnamefont {Piquard}}, \bibinfo {author} {\bibfnamefont {P.}~\bibnamefont {Glidic}}, \bibinfo {author} {\bibfnamefont {C.}~\bibnamefont {Han}}, \bibinfo {author} {\bibfnamefont {A.}~\bibnamefont {Aassime}}, \bibinfo {author} {\bibfnamefont {A.}~\bibnamefont {Cavanna}}, \bibinfo {author} {\bibfnamefont {U.}~\bibnamefont {Gennser}}, \bibinfo {author} {\bibfnamefont {Y.}~\bibnamefont {Meir}}, \bibinfo {author} {\bibfnamefont {E.}~\bibnamefont {Sela}}, \bibinfo {author} {\bibfnamefont {A.}~\bibnamefont {Anthore}},\ and\ \bibinfo {author} {\bibfnamefont {F.}~\bibnamefont {Pierre}},\ }\bibfield  {title} {\bibinfo {title} {Observing the universal screening of a kondo impurity},\ }\href@noop {} {\bibfield  {journal} {\bibinfo  {journal} {Nature Communications}\ }\textbf {\bibinfo {volume} {14}},\ \bibinfo {pages} {7263} (\bibinfo {year} {2023})}\BibitemShut {NoStop}%
\bibitem [{\citenamefont {Child}\ \emph {et~al.}(2022)\citenamefont {Child}, \citenamefont {Sheekey}, \citenamefont {L{\"u}scher}, \citenamefont {Fallahi}, \citenamefont {Gardner}, \citenamefont {Manfra}, \citenamefont {Mitchell}, \citenamefont {Sela}, \citenamefont {Kleeorin}, \citenamefont {Meir} \emph {et~al.}}]{child2022entropy}%
  \BibitemOpen
  \bibfield  {author} {\bibinfo {author} {\bibfnamefont {T.}~\bibnamefont {Child}}, \bibinfo {author} {\bibfnamefont {O.}~\bibnamefont {Sheekey}}, \bibinfo {author} {\bibfnamefont {S.}~\bibnamefont {L{\"u}scher}}, \bibinfo {author} {\bibfnamefont {S.}~\bibnamefont {Fallahi}}, \bibinfo {author} {\bibfnamefont {G.~C.}\ \bibnamefont {Gardner}}, \bibinfo {author} {\bibfnamefont {M.}~\bibnamefont {Manfra}}, \bibinfo {author} {\bibfnamefont {A.}~\bibnamefont {Mitchell}}, \bibinfo {author} {\bibfnamefont {E.}~\bibnamefont {Sela}}, \bibinfo {author} {\bibfnamefont {Y.}~\bibnamefont {Kleeorin}}, \bibinfo {author} {\bibfnamefont {Y.}~\bibnamefont {Meir}}, \emph {et~al.},\ }\bibfield  {title} {\bibinfo {title} {Entropy measurement of a strongly coupled quantum dot},\ }\href@noop {} {\bibfield  {journal} {\bibinfo  {journal} {Phys. Rev. Lett.}\ }\textbf {\bibinfo {volume} {129}},\ \bibinfo {pages} {227702} (\bibinfo {year} {2022})}\BibitemShut {NoStop}%
\bibitem [{\citenamefont {Wilson}(1975)}]{wilson1975renormalization}%
  \BibitemOpen
  \bibfield  {author} {\bibinfo {author} {\bibfnamefont {K.~G.}\ \bibnamefont {Wilson}},\ }\bibfield  {title} {\bibinfo {title} {The renormalization group: Critical phenomena and the kondo problem},\ }\href@noop {} {\bibfield  {journal} {\bibinfo  {journal} {Reviews of modern physics}\ }\textbf {\bibinfo {volume} {47}},\ \bibinfo {pages} {773} (\bibinfo {year} {1975})}\BibitemShut {NoStop}%
\bibitem [{\citenamefont {Bulla}\ \emph {et~al.}(2008)\citenamefont {Bulla}, \citenamefont {Costi},\ and\ \citenamefont {Pruschke}}]{bulla2008numerical}%
  \BibitemOpen
  \bibfield  {author} {\bibinfo {author} {\bibfnamefont {R.}~\bibnamefont {Bulla}}, \bibinfo {author} {\bibfnamefont {T.~A.}\ \bibnamefont {Costi}},\ and\ \bibinfo {author} {\bibfnamefont {T.}~\bibnamefont {Pruschke}},\ }\bibfield  {title} {\bibinfo {title} {Numerical renormalization group method for quantum impurity systems},\ }\href@noop {} {\bibfield  {journal} {\bibinfo  {journal} {Reviews of Modern Physics}\ }\textbf {\bibinfo {volume} {80}},\ \bibinfo {pages} {395} (\bibinfo {year} {2008})}\BibitemShut {NoStop}%
\bibitem [{\citenamefont {Weichselbaum}\ and\ \citenamefont {Von~Delft}(2007)}]{weichselbaum2007sum}%
  \BibitemOpen
  \bibfield  {author} {\bibinfo {author} {\bibfnamefont {A.}~\bibnamefont {Weichselbaum}}\ and\ \bibinfo {author} {\bibfnamefont {J.}~\bibnamefont {Von~Delft}},\ }\bibfield  {title} {\bibinfo {title} {Sum-rule conserving spectral functions from the numerical renormalization group},\ }\href@noop {} {\bibfield  {journal} {\bibinfo  {journal} {Physical review letters}\ }\textbf {\bibinfo {volume} {99}},\ \bibinfo {pages} {076402} (\bibinfo {year} {2007})}\BibitemShut {NoStop}%
\bibitem [{\citenamefont {Mitchell}\ \emph {et~al.}(2014)\citenamefont {Mitchell}, \citenamefont {Galpin}, \citenamefont {Wilson-Fletcher}, \citenamefont {Logan},\ and\ \citenamefont {Bulla}}]{mitchell2014generalized}%
  \BibitemOpen
  \bibfield  {author} {\bibinfo {author} {\bibfnamefont {A.~K.}\ \bibnamefont {Mitchell}}, \bibinfo {author} {\bibfnamefont {M.~R.}\ \bibnamefont {Galpin}}, \bibinfo {author} {\bibfnamefont {S.}~\bibnamefont {Wilson-Fletcher}}, \bibinfo {author} {\bibfnamefont {D.~E.}\ \bibnamefont {Logan}},\ and\ \bibinfo {author} {\bibfnamefont {R.}~\bibnamefont {Bulla}},\ }\bibfield  {title} {\bibinfo {title} {Generalized wilson chain for solving multichannel quantum impurity problems},\ }\href@noop {} {\bibfield  {journal} {\bibinfo  {journal} {Physical Review B}\ }\textbf {\bibinfo {volume} {89}},\ \bibinfo {pages} {121105} (\bibinfo {year} {2014})}\BibitemShut {NoStop}%
\bibitem [{\citenamefont {Stadler}\ \emph {et~al.}(2016)\citenamefont {Stadler}, \citenamefont {Mitchell}, \citenamefont {von Delft},\ and\ \citenamefont {Weichselbaum}}]{stadler2016interleaved}%
  \BibitemOpen
  \bibfield  {author} {\bibinfo {author} {\bibfnamefont {K.}~\bibnamefont {Stadler}}, \bibinfo {author} {\bibfnamefont {A.}~\bibnamefont {Mitchell}}, \bibinfo {author} {\bibfnamefont {J.}~\bibnamefont {von Delft}},\ and\ \bibinfo {author} {\bibfnamefont {A.}~\bibnamefont {Weichselbaum}},\ }\bibfield  {title} {\bibinfo {title} {Interleaved numerical renormalization group as an efficient multiband impurity solver},\ }\href@noop {} {\bibfield  {journal} {\bibinfo  {journal} {Physical Review B}\ }\textbf {\bibinfo {volume} {93}},\ \bibinfo {pages} {235101} (\bibinfo {year} {2016})}\BibitemShut {NoStop}%
\bibitem [{\citenamefont {Rigo}\ and\ \citenamefont {Mitchell}(2022)}]{rigo2022automatic}%
  \BibitemOpen
  \bibfield  {author} {\bibinfo {author} {\bibfnamefont {J.~B.}\ \bibnamefont {Rigo}}\ and\ \bibinfo {author} {\bibfnamefont {A.~K.}\ \bibnamefont {Mitchell}},\ }\bibfield  {title} {\bibinfo {title} {Automatic differentiable numerical renormalization group},\ }\href@noop {} {\bibfield  {journal} {\bibinfo  {journal} {Physical Review Research}\ }\textbf {\bibinfo {volume} {4}},\ \bibinfo {pages} {013227} (\bibinfo {year} {2022})}\BibitemShut {NoStop}%
\bibitem [{\citenamefont {Jarzynski}(2012)}]{jarzynski2012equalities}%
  \BibitemOpen
  \bibfield  {author} {\bibinfo {author} {\bibfnamefont {C.}~\bibnamefont {Jarzynski}},\ }\bibfield  {title} {\bibinfo {title} {Equalities and inequalities: Irreversibility and the second law of thermodynamics at the nanoscale},\ }in\ \href@noop {} {\emph {\bibinfo {booktitle} {Time: Poincar{\'e} Seminar 2010}}}\ (\bibinfo {organization} {Springer},\ \bibinfo {year} {2012})\ pp.\ \bibinfo {pages} {145--172}\BibitemShut {NoStop}%
\bibitem [{\citenamefont {Esposito}\ \emph {et~al.}(2009)\citenamefont {Esposito}, \citenamefont {Harbola},\ and\ \citenamefont {Mukamel}}]{esposito2009nonequilibrium}%
  \BibitemOpen
  \bibfield  {author} {\bibinfo {author} {\bibfnamefont {M.}~\bibnamefont {Esposito}}, \bibinfo {author} {\bibfnamefont {U.}~\bibnamefont {Harbola}},\ and\ \bibinfo {author} {\bibfnamefont {S.}~\bibnamefont {Mukamel}},\ }\bibfield  {title} {\bibinfo {title} {Nonequilibrium fluctuations, fluctuation theorems, and counting statistics in quantum systems},\ }\href@noop {} {\bibfield  {journal} {\bibinfo  {journal} {Rev. Mod. Phys.}\ }\textbf {\bibinfo {volume} {81}},\ \bibinfo {pages} {1665} (\bibinfo {year} {2009})}\BibitemShut {NoStop}%
\bibitem [{\citenamefont {Talkner}\ \emph {et~al.}(2007)\citenamefont {Talkner}, \citenamefont {Lutz},\ and\ \citenamefont {H{\"a}nggi}}]{talkner2007fluctuation}%
  \BibitemOpen
  \bibfield  {author} {\bibinfo {author} {\bibfnamefont {P.}~\bibnamefont {Talkner}}, \bibinfo {author} {\bibfnamefont {E.}~\bibnamefont {Lutz}},\ and\ \bibinfo {author} {\bibfnamefont {P.}~\bibnamefont {H{\"a}nggi}},\ }\bibfield  {title} {\bibinfo {title} {Fluctuation theorems: Work is not an observable},\ }\href@noop {} {\bibfield  {journal} {\bibinfo  {journal} {Phys. Rev. E}\ }\textbf {\bibinfo {volume} {75}},\ \bibinfo {pages} {050102} (\bibinfo {year} {2007})}\BibitemShut {NoStop}%
\bibitem [{\citenamefont {Verley}\ \emph {et~al.}(2014)\citenamefont {Verley}, \citenamefont {Van~den Broeck},\ and\ \citenamefont {Esposito}}]{verley2014work}%
  \BibitemOpen
  \bibfield  {author} {\bibinfo {author} {\bibfnamefont {G.}~\bibnamefont {Verley}}, \bibinfo {author} {\bibfnamefont {C.}~\bibnamefont {Van~den Broeck}},\ and\ \bibinfo {author} {\bibfnamefont {M.}~\bibnamefont {Esposito}},\ }\bibfield  {title} {\bibinfo {title} {Work statistics in stochastically driven systems},\ }\href@noop {} {\bibfield  {journal} {\bibinfo  {journal} {New Journal of Physics}\ }\textbf {\bibinfo {volume} {16}},\ \bibinfo {pages} {095001} (\bibinfo {year} {2014})}\BibitemShut {NoStop}%
\bibitem [{\citenamefont {Landi}\ and\ \citenamefont {Paternostro}(2021)}]{landi2021irreversible}%
  \BibitemOpen
  \bibfield  {author} {\bibinfo {author} {\bibfnamefont {G.~T.}\ \bibnamefont {Landi}}\ and\ \bibinfo {author} {\bibfnamefont {M.}~\bibnamefont {Paternostro}},\ }\bibfield  {title} {\bibinfo {title} {Irreversible entropy production: From classical to quantum},\ }\href@noop {} {\bibfield  {journal} {\bibinfo  {journal} {Rev. Mod. Phys.}\ }\textbf {\bibinfo {volume} {93}},\ \bibinfo {pages} {035008} (\bibinfo {year} {2021})}\BibitemShut {NoStop}%
\bibitem [{\citenamefont {Gherardini}\ and\ \citenamefont {De~Chiara}(2024)}]{gherardini2024quasiprobabilities}%
  \BibitemOpen
  \bibfield  {author} {\bibinfo {author} {\bibfnamefont {S.}~\bibnamefont {Gherardini}}\ and\ \bibinfo {author} {\bibfnamefont {G.}~\bibnamefont {De~Chiara}},\ }\bibfield  {title} {\bibinfo {title} {Quasiprobabilities in quantum thermodynamics and many-body systems},\ }\href@noop {} {\bibfield  {journal} {\bibinfo  {journal} {{PRX} {Q}uantum}\ }\textbf {\bibinfo {volume} {5}},\ \bibinfo {pages} {030201} (\bibinfo {year} {2024})}\BibitemShut {NoStop}%
\bibitem [{\citenamefont {Strasberg}\ and\ \citenamefont {Winter}(2021)}]{strasberg2021first}%
  \BibitemOpen
  \bibfield  {author} {\bibinfo {author} {\bibfnamefont {P.}~\bibnamefont {Strasberg}}\ and\ \bibinfo {author} {\bibfnamefont {A.}~\bibnamefont {Winter}},\ }\bibfield  {title} {\bibinfo {title} {First and second law of quantum thermodynamics: A consistent derivation based on a microscopic definition of entropy},\ }\href@noop {} {\bibfield  {journal} {\bibinfo  {journal} {{PRX} {Q}uantum}\ }\textbf {\bibinfo {volume} {2}},\ \bibinfo {pages} {030202} (\bibinfo {year} {2021})}\BibitemShut {NoStop}%
\bibitem [{SM()}]{SM}%
  \BibitemOpen
  \href@noop {} {}\bibinfo {note} {See Supplemental Material at http://link.aps.org/supplemental/10.1103/vn83-mt2v, which includes Ref.[47,48,52], for further details on (i) the derivation of Eq.3, Eq.5, Eq.6 and Eq.10. (ii) the physical condition for $c(T)=0$ in Eq.5 (iii) the noninteracting QD case}\BibitemShut {NoStop}%
\bibitem [{\citenamefont {Kubo}(1957)}]{kubo1957statistical}%
  \BibitemOpen
  \bibfield  {author} {\bibinfo {author} {\bibfnamefont {R.}~\bibnamefont {Kubo}},\ }\bibfield  {title} {\bibinfo {title} {Statistical-mechanical theory of irreversible processes. i. general theory and simple applications to magnetic and conduction problems},\ }\href@noop {} {\bibfield  {journal} {\bibinfo  {journal} {Journal of the physical society of Japan}\ }\textbf {\bibinfo {volume} {12}},\ \bibinfo {pages} {570} (\bibinfo {year} {1957})}\BibitemShut {NoStop}%
\bibitem [{\citenamefont {Filippone}(2013)}]{filippone:tel-00908428}%
  \BibitemOpen
  \bibfield  {author} {\bibinfo {author} {\bibfnamefont {M.}~\bibnamefont {Filippone}},\ }\emph {\bibinfo {title} {Fermi liquid theory of the strongly interacting quantum {RC} circuit}},\ \href {https://theses.hal.science/tel-00908428} {\bibinfo {type} {Theses}},\ \bibinfo  {school} {{Ecole Normale Sup{\'e}rieure de Paris - ENS Paris}} (\bibinfo {year} {2013})\BibitemShut {NoStop}%
\bibitem [{\citenamefont {Francica}\ \emph {et~al.}(2019)\citenamefont {Francica}, \citenamefont {Goold},\ and\ \citenamefont {Plastina}}]{francica2019role}%
  \BibitemOpen
  \bibfield  {author} {\bibinfo {author} {\bibfnamefont {G.}~\bibnamefont {Francica}}, \bibinfo {author} {\bibfnamefont {J.}~\bibnamefont {Goold}},\ and\ \bibinfo {author} {\bibfnamefont {F.}~\bibnamefont {Plastina}},\ }\bibfield  {title} {\bibinfo {title} {Role of coherence in the nonequilibrium thermodynamics of quantum systems},\ }\href@noop {} {\bibfield  {journal} {\bibinfo  {journal} {Physical Review E}\ }\textbf {\bibinfo {volume} {99}},\ \bibinfo {pages} {042105} (\bibinfo {year} {2019})}\BibitemShut {NoStop}%
\bibitem [{\citenamefont {Santos}\ \emph {et~al.}(2019)\citenamefont {Santos}, \citenamefont {C{\'e}leri}, \citenamefont {Landi},\ and\ \citenamefont {Paternostro}}]{santos2019role}%
  \BibitemOpen
  \bibfield  {author} {\bibinfo {author} {\bibfnamefont {J.~P.}\ \bibnamefont {Santos}}, \bibinfo {author} {\bibfnamefont {L.~C.}\ \bibnamefont {C{\'e}leri}}, \bibinfo {author} {\bibfnamefont {G.~T.}\ \bibnamefont {Landi}},\ and\ \bibinfo {author} {\bibfnamefont {M.}~\bibnamefont {Paternostro}},\ }\bibfield  {title} {\bibinfo {title} {The role of quantum coherence in non-equilibrium entropy production},\ }\href@noop {} {\bibfield  {journal} {\bibinfo  {journal} {npj Quantum Information}\ }\textbf {\bibinfo {volume} {5}},\ \bibinfo {pages} {23} (\bibinfo {year} {2019})}\BibitemShut {NoStop}%
\bibitem [{\citenamefont {Kiely}\ \emph {et~al.}(2023)\citenamefont {Kiely}, \citenamefont {O'Connor}, \citenamefont {Fogarty}, \citenamefont {Landi},\ and\ \citenamefont {Campbell}}]{kiely2023entropy}%
  \BibitemOpen
  \bibfield  {author} {\bibinfo {author} {\bibfnamefont {A.}~\bibnamefont {Kiely}}, \bibinfo {author} {\bibfnamefont {E.}~\bibnamefont {O'Connor}}, \bibinfo {author} {\bibfnamefont {T.}~\bibnamefont {Fogarty}}, \bibinfo {author} {\bibfnamefont {G.~T.}\ \bibnamefont {Landi}},\ and\ \bibinfo {author} {\bibfnamefont {S.}~\bibnamefont {Campbell}},\ }\bibfield  {title} {\bibinfo {title} {Entropy of the quantum work distribution},\ }\href@noop {} {\bibfield  {journal} {\bibinfo  {journal} {Physical Review Research}\ }\textbf {\bibinfo {volume} {5}},\ \bibinfo {pages} {L022010} (\bibinfo {year} {2023})}\BibitemShut {NoStop}%
\bibitem [{\citenamefont {Guarnieri}\ \emph {et~al.}(2024)\citenamefont {Guarnieri}, \citenamefont {Eisert},\ and\ \citenamefont {Miller}}]{guarnieri2023generalised}%
  \BibitemOpen
  \bibfield  {author} {\bibinfo {author} {\bibfnamefont {G.}~\bibnamefont {Guarnieri}}, \bibinfo {author} {\bibfnamefont {J.}~\bibnamefont {Eisert}},\ and\ \bibinfo {author} {\bibfnamefont {H.~J.}\ \bibnamefont {Miller}},\ }\bibfield  {title} {\bibinfo {title} {Generalized linear response theory for the full quantum work statistics},\ }\href@noop {} {\bibfield  {journal} {\bibinfo  {journal} {Physical Review Letters}\ }\textbf {\bibinfo {volume} {133}},\ \bibinfo {pages} {070405} (\bibinfo {year} {2024})}\BibitemShut {NoStop}%
\bibitem [{\citenamefont {Furusaki}\ and\ \citenamefont {Matveev}(1995)}]{furusaki1995theory}%
  \BibitemOpen
  \bibfield  {author} {\bibinfo {author} {\bibfnamefont {A.}~\bibnamefont {Furusaki}}\ and\ \bibinfo {author} {\bibfnamefont {K.}~\bibnamefont {Matveev}},\ }\bibfield  {title} {\bibinfo {title} {Theory of strong inelastic cotunneling},\ }\href@noop {} {\bibfield  {journal} {\bibinfo  {journal} {Physical Review B}\ }\textbf {\bibinfo {volume} {52}},\ \bibinfo {pages} {16676} (\bibinfo {year} {1995})}\BibitemShut {NoStop}%
\bibitem [{\citenamefont {Affleck}\ and\ \citenamefont {Ludwig}(1993)}]{affleck1993exact}%
  \BibitemOpen
  \bibfield  {author} {\bibinfo {author} {\bibfnamefont {I.}~\bibnamefont {Affleck}}\ and\ \bibinfo {author} {\bibfnamefont {A.~W.}\ \bibnamefont {Ludwig}},\ }\bibfield  {title} {\bibinfo {title} {Exact conformal-field-theory results on the multichannel kondo effect: Single-fermion green’s function, self-energy, and resistivity},\ }\href@noop {} {\bibfield  {journal} {\bibinfo  {journal} {Physical Review B}\ }\textbf {\bibinfo {volume} {48}},\ \bibinfo {pages} {7297} (\bibinfo {year} {1993})}\BibitemShut {NoStop}%
\bibitem [{\citenamefont {Emery}\ and\ \citenamefont {Kivelson}(1992)}]{emery1992mapping}%
  \BibitemOpen
  \bibfield  {author} {\bibinfo {author} {\bibfnamefont {V.}~\bibnamefont {Emery}}\ and\ \bibinfo {author} {\bibfnamefont {S.}~\bibnamefont {Kivelson}},\ }\bibfield  {title} {\bibinfo {title} {Mapping of the two-channel kondo problem to a resonant-level model},\ }\href@noop {} {\bibfield  {journal} {\bibinfo  {journal} {Physical Review B}\ }\textbf {\bibinfo {volume} {46}},\ \bibinfo {pages} {10812} (\bibinfo {year} {1992})}\BibitemShut {NoStop}%
\bibitem [{\citenamefont {Lopes}\ \emph {et~al.}(2020)\citenamefont {Lopes}, \citenamefont {Affleck},\ and\ \citenamefont {Sela}}]{lopes2020anyons}%
  \BibitemOpen
  \bibfield  {author} {\bibinfo {author} {\bibfnamefont {P.~L.}\ \bibnamefont {Lopes}}, \bibinfo {author} {\bibfnamefont {I.}~\bibnamefont {Affleck}},\ and\ \bibinfo {author} {\bibfnamefont {E.}~\bibnamefont {Sela}},\ }\bibfield  {title} {\bibinfo {title} {Anyons in multichannel kondo systems},\ }\href@noop {} {\bibfield  {journal} {\bibinfo  {journal} {Physical Review B}\ }\textbf {\bibinfo {volume} {101}},\ \bibinfo {pages} {085141} (\bibinfo {year} {2020})}\BibitemShut {NoStop}%
\bibitem [{\citenamefont {Lotem}\ \emph {et~al.}(2022)\citenamefont {Lotem}, \citenamefont {Sela},\ and\ \citenamefont {Goldstein}}]{PhysRevLett.129.227703}%
  \BibitemOpen
  \bibfield  {author} {\bibinfo {author} {\bibfnamefont {M.}~\bibnamefont {Lotem}}, \bibinfo {author} {\bibfnamefont {E.}~\bibnamefont {Sela}},\ and\ \bibinfo {author} {\bibfnamefont {M.}~\bibnamefont {Goldstein}},\ }\bibfield  {title} {\bibinfo {title} {Manipulating non-abelian anyons in a chiral multichannel kondo model},\ }\href {https://doi.org/10.1103/PhysRevLett.129.227703} {\bibfield  {journal} {\bibinfo  {journal} {Phys. Rev. Lett.}\ }\textbf {\bibinfo {volume} {129}},\ \bibinfo {pages} {227703} (\bibinfo {year} {2022})}\BibitemShut {NoStop}%
\bibitem [{\citenamefont {Han}\ \emph {et~al.}(2024)\citenamefont {Han}, \citenamefont {Cohen},\ and\ \citenamefont {Sela}}]{PhysRevB.110.115153}%
  \BibitemOpen
  \bibfield  {author} {\bibinfo {author} {\bibfnamefont {C.}~\bibnamefont {Han}}, \bibinfo {author} {\bibfnamefont {D.}~\bibnamefont {Cohen}},\ and\ \bibinfo {author} {\bibfnamefont {E.}~\bibnamefont {Sela}},\ }\bibfield  {title} {\bibinfo {title} {Quantum limitation on experimental testing of nonequilibrium fluctuation theorems},\ }\href@noop {} {\bibfield  {journal} {\bibinfo  {journal} {Phys. Rev. B}\ }\textbf {\bibinfo {volume} {110}},\ \bibinfo {pages} {115153} (\bibinfo {year} {2024})}\BibitemShut {NoStop}%
\bibitem [{\citenamefont {Mihailescu}\ \emph {et~al.}(2024)\citenamefont {Mihailescu}, \citenamefont {Kiely},\ and\ \citenamefont {Mitchell}}]{mihailescu2024quantum}%
  \BibitemOpen
  \bibfield  {author} {\bibinfo {author} {\bibfnamefont {G.}~\bibnamefont {Mihailescu}}, \bibinfo {author} {\bibfnamefont {A.}~\bibnamefont {Kiely}},\ and\ \bibinfo {author} {\bibfnamefont {A.~K.}\ \bibnamefont {Mitchell}},\ }\bibfield  {title} {\bibinfo {title} {Quantum sensing with nanoelectronics: {F}isher information for a perturbation},\ }\href@noop {} {\bibfield  {journal} {\bibinfo  {journal} {ar{X}iv preprint ar{X}iv:2406.18662}\ } (\bibinfo {year} {2024})}\BibitemShut {NoStop}%
\bibitem [{\citenamefont {Grabarits}\ \emph {et~al.}(2025)\citenamefont {Grabarits}, \citenamefont {Balducci},\ and\ \citenamefont {del Campo}}]{PhysRevA.111.042207}%
  \BibitemOpen
  \bibfield  {author} {\bibinfo {author} {\bibfnamefont {A.}~\bibnamefont {Grabarits}}, \bibinfo {author} {\bibfnamefont {F.}~\bibnamefont {Balducci}},\ and\ \bibinfo {author} {\bibfnamefont {A.}~\bibnamefont {del Campo}},\ }\bibfield  {title} {\bibinfo {title} {Driving a quantum phase transition at an arbitrary rate: Exact solution of the transverse-field ising model},\ }\href {https://doi.org/10.1103/PhysRevA.111.042207} {\bibfield  {journal} {\bibinfo  {journal} {Phys. Rev. A}\ }\textbf {\bibinfo {volume} {111}},\ \bibinfo {pages} {042207} (\bibinfo {year} {2025})}\BibitemShut {NoStop}%
\end{thebibliography}%


\begin{thebibliography}{3}%
\makeatletter
\providecommand \@ifxundefined [1]{%
 \@ifx{#1\undefined}
}%
\providecommand \@ifnum [1]{%
 \ifnum #1\expandafter \@firstoftwo
 \else \expandafter \@secondoftwo
 \fi
}%
\providecommand \@ifx [1]{%
 \ifx #1\expandafter \@firstoftwo
 \else \expandafter \@secondoftwo
 \fi
}%
\providecommand \natexlab [1]{#1}%
\providecommand \enquote  [1]{``#1''}%
\providecommand \bibnamefont  [1]{#1}%
\providecommand \bibfnamefont [1]{#1}%
\providecommand \citenamefont [1]{#1}%
\providecommand \href@noop [0]{\@secondoftwo}%
\providecommand \href [0]{\begingroup \@sanitize@url \@href}%
\providecommand \@href[1]{\@@startlink{#1}\@@href}%
\providecommand \@@href[1]{\endgroup#1\@@endlink}%
\providecommand \@sanitize@url [0]{\catcode `\\12\catcode `\$12\catcode `\&12\catcode `\#12\catcode `\^12\catcode `\_12\catcode `\%12\relax}%
\providecommand \@@startlink[1]{}%
\providecommand \@@endlink[0]{}%
\providecommand \url  [0]{\begingroup\@sanitize@url \@url }%
\providecommand \@url [1]{\endgroup\@href {#1}{\urlprefix }}%
\providecommand \urlprefix  [0]{URL }%
\providecommand \Eprint [0]{\href }%
\providecommand \doibase [0]{http://dx.doi.org/}%
\providecommand \selectlanguage [0]{\@gobble}%
\providecommand \bibinfo  [0]{\@secondoftwo}%
\providecommand \bibfield  [0]{\@secondoftwo}%
\providecommand \translation [1]{[#1]}%
\providecommand \BibitemOpen [0]{}%
\providecommand \bibitemStop [0]{}%
\providecommand \bibitemNoStop [0]{.\EOS\space}%
\providecommand \EOS [0]{\spacefactor3000\relax}%
\providecommand \BibitemShut  [1]{\csname bibitem#1\endcsname}%
\let\auto@bib@innerbib\@empty
%</preamble>
\bibitem [{\citenamefont {Kubo}(1957)}]{kubo1957statistical}%
  \BibitemOpen
  \bibfield  {author} {\bibinfo {author} {\bibfnamefont {Ryogo}\ \bibnamefont {Kubo}},\ }\bibfield  {title} {\enquote {\bibinfo {title} {Statistical-mechanical theory of irreversible processes. i. general theory and simple applications to magnetic and conduction problems},}\ }\href@noop {} {\bibfield  {journal} {\bibinfo  {journal} {Journal of the physical society of Japan}\ }\textbf {\bibinfo {volume} {12}},\ \bibinfo {pages} {570--586} (\bibinfo {year} {1957})}\BibitemShut {NoStop}%
\bibitem [{\citenamefont {Guarnieri}\ \emph {et~al.}(2024)\citenamefont {Guarnieri}, \citenamefont {Eisert},\ and\ \citenamefont {Miller}}]{guarnieri2023generalised}%
  \BibitemOpen
  \bibfield  {author} {\bibinfo {author} {\bibfnamefont {Giacomo}\ \bibnamefont {Guarnieri}}, \bibinfo {author} {\bibfnamefont {Jens}\ \bibnamefont {Eisert}}, \ and\ \bibinfo {author} {\bibfnamefont {Harry~JD}\ \bibnamefont {Miller}},\ }\bibfield  {title} {\enquote {\bibinfo {title} {Generalized linear response theory for the full quantum work statistics},}\ }\href@noop {} {\bibfield  {journal} {\bibinfo  {journal} {Physical Review Letters}\ }\textbf {\bibinfo {volume} {133}},\ \bibinfo {pages} {070405} (\bibinfo {year} {2024})}\BibitemShut {NoStop}%
\bibitem [{\citenamefont {Filippone}(2013)}]{filippone:tel-00908428}%
  \BibitemOpen
  \bibfield  {author} {\bibinfo {author} {\bibfnamefont {Michele}\ \bibnamefont {Filippone}},\ }\emph {\bibinfo {title} {Fermi liquid theory of the strongly interacting quantum {RC} circuit}},\ \href {https://theses.hal.science/tel-00908428} {\bibinfo {type} {Theses}},\ \bibinfo  {school} {{Ecole Normale Sup{\'e}rieure de Paris - ENS Paris}} (\bibinfo {year} {2013})\BibitemShut {NoStop}%
\end{thebibliography}%

\end{document}

\title{\textbf{\underline{SUPPLEMENTARY MATERIAL}}\\Quantum work statistics across a critical point:\\
full crossover from sudden quench to the adiabatic limit}

\author{Zhanyu Ma}
\affiliation{Raymond and Beverly Sackler School of Physics and Astronomy, Tel Aviv University, Tel Aviv 69978, Israel} 
\author{Andrew K. Mitchell}
\affiliation{School of Physics, University College Dublin, Belfield, Dublin 4, Ireland} 
\affiliation{Centre for Quantum Engineering, Science, and Technology, University College Dublin, Dublin, Ireland}
\author{Eran Sela}
\affiliation{Raymond and Beverly Sackler School of Physics and Astronomy, Tel Aviv University, Tel Aviv 69978, Israel}

\maketitle
%%%%%%%%%%%%%%%%%%%%%%%%%%%%%%%%%%%%%%%%%%%%%%%%%%%%%%%%%%%%%%%%%%%%%%%
%%%%%%%%%%%%%%%%%%%%%%%%%%%%%%%%%%%%%%%%%%%%%%%%%%%%%%%%%%%%%%%%%%%%%%%

\section{Derivation of Eq.~3}
In the sudden-quench limit, $U_{\tau}\to \gi$, the generating function $h(u)$, given in Eq.~2 in the main text, reduces to
\be
\label{eq:generating-sudden}
h(u)={\rm{Tr}} [ e^{-u (\hat{H}_0+A\hat{N}_{\mathcal{S}})} e^{-(\beta-u)\hat{H}_0} ]/\mathcal{Z}\;.
\ee
Here $\mathcal{Z}$ is the partition function of $\hat{H}_0$ and $\beta=1/T$ (with $k_B,\hbar \equiv 1$).
Using the Zassenhaus formula, the first term inside the trace $e^{-u (\hat{H}_0+A\hat{N}_{\mathcal{S}})}$ can be expanded as an infinite product
\be
e^{-u (\hat{H}_0+A\hat{N}_{\mathcal{S}})}=e^{-u \hat{H}_0}e^{-u A \hat{N}_{\mathcal{S}}}e^{-\frac{u^2}{2} [\hat{H}_0,A\hat{N}_{\mathcal{S}}]}
e^{-\frac{u^3}{6}(2[A\hat{N}_{\mathcal{S}}, [\hat{H}_0, A\hat{N}_{\mathcal{S}}]] + [\hat{H}_0, [\hat{H}_0, A\hat{N}_{\mathcal{S}}]])}\cdots,
\label{eq:za_exp}
\ee
where $\cdots$ represents exponents of higher order in $u$. Substituting Eq.~\ref{eq:za_exp} into Eq.~\ref{eq:generating-sudden}, 
we have
\be
\label{eq:zassenhaus}
h(u)=\langle  e^{-u A \hat{N}_{\mathcal{S}}}e^{-\frac{u^2}{2} [\hat{H}_0,A\hat{N}_{\mathcal{S}}]}
e^{-\frac{u^3}{6}(2[A\hat{N}_{\mathcal{S}}, [\hat{H}_0, A\hat{N}_{\mathcal{S}}]] + [\hat{H}_0, [\hat{H}_0, A\hat{N}_{\mathcal{S}}]])}\cdots
\rangle_0,
\ee
where $\langle \cdot \rangle_0={\rm{Tr}} \{  \rho_0 \cdot \}$. The $n^{\rm th}$ moment of the WDF is given by $\langle W^n \rangle = (-1)^n \frac{d^n}{du^n} h(u)|_{u=0}$, and thus can be calculated from Eq.~\ref{eq:zassenhaus}
by keeping the first $n$ terms. The classical contribution comes from the first exponent: Expanding $e^{-u A \hat{N}_{\mathcal{S}}}$ as a series of powers in $u$ and substituting it into $\langle W^n \rangle = (-1)^n \frac{d^n}{du^n} h(u)|_{u=0}$ results in the term given by $A^n\langle\hat{N}_{\mathcal{S}}^n\rangle_0$ in Eq.~3. All other contributions originating from higher exponents are defined as $\delta \mathcal{Q}_n$, and, in particular, we find
$\delta \mathcal{Q}_1=\delta \mathcal{Q}_2=0$, $\delta \mathcal{Q}_3=\frac{A^2}{2} \langle [\hat{N}_{\mathcal{S}},[\hat{H}_0,\hat{N}_{\mathcal{S}}]] \rangle_0$.

%%%%%%%%%%%%%%%%%

\section{Derivation of Eq.~5}
Our derivation is made up of 2 steps. First, we show that $\frac{1}{\beta}\frac{d\Psi_0(t)}{dt}=-i{\rm{Tr}}\left[\rho [\hat{N}_{\mathcal{S}}(t), \hat{N}_{\mathcal{S}}(0)]\right]$, following Ref.~\onlinecite{kubo1957statistical}. 
Recall the definition of $\Psi_0(t)$, 
\be
\Psi_0(t)=\beta \int_0^\beta ds \: \langle \hat{N}_{\mathcal{S}}(-is) \hat{N}_{\mathcal{S}}(t) \rangle_0 -\beta^2 \langle \hat{N}_{\mathcal{S}}\rangle_0^2 \;,
\ee
where $\hat{N}_{\mathcal{S}}(-is) = e^{s\hat{H}_0}\hat{N}_{\mathcal{S}}e^{-s\hat{H}_0}$, and $\hat{N}_{\mathcal{S}}(t) = e^{i\hat{H}_0t}\hat{N}_{\mathcal{S}}e^{-i\hat{H}_0t}$. Using the identity
\be
\int_0^{\beta}e^{-\lambda \hat{H}_0}[\hat{H}_0, \hat{A}]e^{\lambda \hat{H}_0}d\lambda = -e^{-\beta \hat{H}_0}[\hat{A}, e^{\beta \hat{H}_0}],
\ee
which holds for any operator $\hat{A}$,
we have
\begin{align}
\frac{1}{\beta}\frac{d\Psi_0(t)}{dt} &= i\int_0^{\beta}ds\langle \hat{N}_{\mathcal{S}}(-is) [\hat{H}_0, \hat{N}_{\mathcal{S}}(t)] \rangle\nonumber\\
&=i\int_0^{\beta}ds{\rm{Tr}}\left[{\frac{e^{-\beta \hat{H}_0}}{\mathcal{Z}}e^{s\hat{H}_0}\hat{N}_{\mathcal{S}}e^{-s\hat{H}_0}[\hat{H}_0, \hat{N}_{\mathcal{S}}(t)]}\right]\nonumber\\
&=i\int_0^{\beta}ds{\rm{Tr}}\left[{\frac{e^{-\beta \hat{H}_0}}{\mathcal{Z}}\hat{N}_{\mathcal{S}}e^{-s\hat{H}_0}[\hat{H}_0, \hat{N}_{\mathcal{S}}(t)]e^{s\hat{H}_0}}\right]\nonumber\\
&=-i{\rm{Tr}}\left[{\frac{e^{-\beta \hat{H}_0}}{\mathcal{Z}}\hat{N}_{\mathcal{S}}e^{-\beta \hat{H}_0}[\hat{N}_{\mathcal{S}}(t), e^{\beta \hat{H}_0}]}\right]\nonumber\\
&= -i{\rm{Tr}}\left[{\rho_0 [\hat{N}_{\mathcal{S}}(t), \hat{N}_{\mathcal{S}}]}\right].
\label{eq:Psi_dt}
\end{align}

Second, we express the Fourier transform of $f(t)=-i{\rm{Tr}}\left[\rho [\hat{N}_{\mathcal{S}}(t), \hat{N}_{\mathcal{S}}(0)]\right]$ in terms of the Fourier transform of $\chi(t)=i\theta(t){\rm Tr}\left[\rho_0 [\hat{N}_{\mathcal{S}}(t),\hat{N}_{\mathcal{S}}]\right]$. By definition, $\chi(t)=-\theta(t)f(t)$, thus we have
\begin{align*}
\chi(\omega) &= -\int_0^{\infty} dt f(t) e^{i\omega t - \eta t}\\
& = -\int_{-\infty}^{\infty} \frac{d\Omega}{2\pi} f(\Omega)\int_0^{\infty}dt e^{i(\omega-\Omega)t-\eta t}\\
& = -\int_{-\infty}^{\infty} \frac{d\Omega}{2\pi} f(\Omega) \frac{-i}{\Omega-\omega-i\eta}\\
& = \mathcal{P}\int\frac{d\Omega}{2\pi} if(\Omega) \frac{1}{\Omega - \omega} - \frac{1}{2} f(\omega).
\end{align*}
Now, since $f(t)=-f(-t)$ it follows that $f(\omega)$ is purely imaginary, so that the first term in the last line is purely real. We can eliminate it by taking the imaginary part of the equation, which yields $f(\omega) = -2i{\rm{Im}} \chi(\omega)$. 
Finally, noticing that $\frac{1}{\beta}\frac{d\Psi_0(t)}{dt}=f(t)$, we conclude
\be
\tilde{\Psi}_0(\omega) = \frac{2}{\omega T}{\rm{Im}}\chi(\omega) +2\pi\frac{c(T)}{T}\delta(\omega). 
\label{eq:Psi_fre}
\ee
The second term proportional to $\delta(\omega)$ could in principle appear if  the relaxation function $\Psi_0(t)$ contains a finite constant term (that is, $\Psi_0(t)\to c(T)/T$  as $|t|\to \infty$).  

%%%%%
\subsection{$c(T)$ from  Lehmann representation}
Below we give an expression for $c(T)$ using a Lehmann representation. We thank Mark Mitchison for pointing this out to us.
\zm{First, we notice that
\be
\langle \hat{N}_{\mathcal{S}}(-is)\hat{N}_{\mathcal{S}}(t) \rangle_0 = \sum_{m,n}\frac{e^{-\beta E_m}}{\mathcal{Z}}e^{-(s-it)(E_n-E_m)}|\langle m | \hat{N}_{\mathcal{S}} | n\rangle|^2.
\ee
To integrate over $s$, we separate the sum into terms with $E_m=E_n$ and terms with $E_m \ne E_n$, and we get
\be
\int_0^{\beta}ds \langle \hat{N}_{\mathcal{S}}(-is)\hat{N}_{\mathcal{S}}(t) \rangle_0 = \sum_{E_m\neq E_n} \frac{1}{E_m-E_n}\frac{e^{-\beta E_n}-e^{-\beta E_m}}{\mathcal{Z}}  e^{it(E_n - E_m)}|\langle m|\hat{N}_{\mathcal{S}}|n\rangle|^2+\beta\sum_{E_m=E_n} \frac{e^{-\beta E_m}}{\mathcal{Z}}|\langle m|\hat{N}_{\mathcal{S}}|n\rangle|^2.
\ee
The terms with $E_m=E_n$, together with the term $\beta^2\langle\hat{N}_{\mathcal{S}}\rangle_0^2$, give the $c(T)$ term in Eq.~\ref{eq:Psi_fre} upon Fourier transformation.
The Lehmann representation of the relaxation function shows that
\be
\label{eq:Lehman_c_T}
c(T) = \beta \left(\sum_{E_m=E_n}\frac{e^{-\beta E_m}}{\mathcal{Z}}|\langle m | \hat{N}_{\mathcal{S}} |n \rangle|^2 - \langle \hat{N}_{\mathcal{S}} \rangle_0^2\right).
\ee
}

%%%%%%%%%

\section{Physical condition for $c(T)=0$}
\label{se:ceqz}
We can obtain a relation of $c(T)$ from Eq.~\ref{eq:Psi_fre} by integration over $\omega$, which gives the relaxation function at $t=0$. As given in Eq.~9 in the main text, and derived in Sec.~\ref{se:9_10},  this gives the static susceptibility,
\be
-\frac{d\langle\hat{N}_{\mathcal{S}}\rangle}{d\lambda} = T\int \frac{d\omega}{2\pi} \Psi_0(\omega)=\frac{1}{\pi}\int d\omega \frac{{\rm{Im}}\chi(\omega)}{\omega} + c(T),
\ee
thus we have 
\be
c(T)=-\frac{ d\langle\hat{N}_{\mathcal{S}}\rangle}{d\lambda}  - \frac{1}{\pi}\int d\omega \frac{{\rm{Im}}\chi(\omega)}{\omega} \;,
\label{eq:coft}
\ee
where the static susceptibility $\frac{ d\langle\hat{N}_{\mathcal{S}}\rangle}{d\lambda}$ is evaluated at $\lambda=0$. 
Since the retarded correlation function $\chi(\omega)$ is analytic in the upper-half complex plane, the Kramers-Kronig relation yields,
\be
{\rm{Re}\chi(\omega=0)} = \frac{1}{\pi}\int d\omega \frac{{\rm{Im}}\chi(\omega)}{\omega}.
\ee
This gives
\be\label{eq:SI_c}
c(T)=-\frac{d\langle\hat{N}_{\mathcal{S}}\rangle}{d\lambda} -{\rm{Re}\chi(\omega=0)}
\ee
At a given temperature $T$, typically one can deduce that ${\rm{Re}\chi(\omega=0)}=-\frac{d\langle\hat{N}_{\mathcal{S}}\rangle_0}{d\lambda}$ according to linear response theory, so that $c(T)=0$.  This indeed holds in all the cases studied in our work, where we have open quantum systems and an environment with a continuous, gapless spectrum. The condition for $c(T)=0$ is therefore that the static susceptibility $\frac{d\langle\hat{N}_{\mathcal{S}}\rangle}{d\lambda}$ can be obtained from the Kubo formula in the usual way. This can break down for closed finite systems or open systems with discrete environments.

%%%%%%%

\subsection{Explicit demonstration that $c(T)=0$ in RLM and Majorana RLM}
For the RLM and Majorana RLM, we have analytic expressions for both $\chi(T)=\frac{d\langle\hat{N}_{\mathcal{S}}\rangle}{d\lambda}$ and ${\rm{Im}}\chi(\omega)$, so $c(T)$ can be extracted explicitly. We confirm that $c(T)=0$ in both cases.

For Majorana RLM, we have 
\be
    {\rm{Im}}\chi(\omega) = \frac{1}{2}\tanh{\left(\frac{\omega}{2T}\right)}\frac{\Gamma}{\omega^2+\Gamma^2},
    \label{eq:mrlm_chi}
\ee
at $\lambda=0$ and 
\be
   \langle \hat{N}_{\mathcal{S}} \rangle_0 = \frac{1}{2} - {\rm{Im}}\left\{\frac{\lambda}{\pi\Delta}\left[\psi(\frac{1}{2} + \frac{\Gamma + i\Delta}{4\pi T}) - \psi(\frac{1}{2} + \frac{\Gamma - i\Delta}{4\pi T})\right]\right\},
\ee
where $\Delta=\sqrt{4\lambda^2-\Gamma^2}$. Substituting into Eq.~\ref{eq:SI_c} we indeed get $c(T)=0$.

Similarly, for RLM, we have
\be
{\rm{Im}}\chi(\omega) = {\rm{Im}}\left[\frac{2}{\pi\Gamma}\frac{1}{\frac{\omega^2}{\Gamma^2}+2i\frac{\omega}{\Gamma}}\times\right. \\
  \left.\left(\psi(\frac{1}{2} + \frac{\Gamma}{2\pi T}) - \psi(\frac{1}{2} + \frac{\Gamma}{2\pi T} - i\frac{\omega}{2\pi T})\right)\right],
\ee
and 
\be
\langle \hat{N}_{\mathcal{S}} \rangle_0 = \frac{1}{2} + \frac{1}{2\pi}{\rm{Im}}\left[\psi(\frac{1}{2} + \frac{\Gamma - i\lambda}{2\pi T}) - \psi(\frac{1}{2} + \frac{\Gamma + i\lambda}{2\pi T}) \right].
\ee
Again we find $c(T)=0$ via Eq.~\ref{eq:SI_c}.

Let us point out that we have checked that $c(T)=0$ also for the C2CK and C3CK models using our NRG calculations.

%%%%%%%%%%

\subsection{Example of a scenario where $c(T) \ne 0$}
In situations where the Kubo formula for the static susceptibility $\frac{d\langle\hat{N}_{\mathcal{S}}\rangle}{d\lambda}$ in terms of the dynamical susceptibility $\chi(\omega)$ would not apply, one might expect that $c(T)$ could be finite. To illustrate this scenario,
consider the following simple Hamiltonian $\hat{H}_0$ for an isolated two-level system, and the operator $\hat{N}_{\mathcal{S}}$,
\be
\hat{H}_0 = w
\begin{pmatrix}
x & 1\\
1 & 0
\end{pmatrix},
\quad
\hat{N}_{\mathcal{S}}=
\begin{pmatrix}
1 & 0\\
0 & 0
\end{pmatrix}.
\ee
The retarded  correlator $\chi(t)=i\theta(t){\rm{Tr}}\left\{\rho_0\left[\hat{N}_{\mathcal{S}}(t), \hat{N}_{\mathcal{S}}(0)\right]\right\}$ is in this case given by,
\be
\chi(t)=\frac{2\tanh\left(\frac{\beta w}{2}\sqrt{4+x^2}\right)}{4+x^2}\sin\left(wt\sqrt{4+x^2}\right)\theta(t),
\ee
 from which we find,
\be
{\rm{Re}}\chi(\omega=0)=\frac{2\tanh\left(\frac{\beta w}{2}\sqrt{4+x^2}\right)}{w(4+x^2)^{3/2}}.
\label{eq:dd_sus}
\ee
On the other hand, a direct calculation of the susceptibility gives
\be
\frac{d\langle\hat{N}_{\mathcal{S}}\rangle}{d\lambda} = -\frac{1}{w} \frac{\beta w x^2/(4+x^2) + 4\sinh\left(\beta w\sqrt{4+x^2}\right)/(4+x^2)^{3/2}}{2\left(1+\cosh\left(\beta w \sqrt{4+x^2}\right)\right)} \;.
\label{eq:dd_dndt}
\ee
Substituting Eq.~(\ref{eq:dd_sus}) and Eq.~(\ref{eq:dd_dndt}) into Eq.~(\ref{eq:coft}), we get
\be
c(T) = \beta \frac{x^2}{2(4+x^2)\left(1+\cosh\left(\beta w \sqrt{4+x^2}\right)\right)} \;,
\label{eq:dd_coft}
\ee
which is indeed finite for any finite $x$ and $T$. The Lehmann representation of the relaxation function, Eq.~\ref{eq:Lehman_c_T} can be checked to match Eq.~\ref{eq:dd_coft} exactly.

%%%%%%%%%%

\subsection{Discussion}
The only possibility for the non-cancellation of the two terms contributing to Eq.~\ref{eq:coft} is that the time evolution of the unperturbed state under the Hamiltonian $H(t)=H_0 + \delta \lambda e^{ \delta t} \hat{N}_\mathcal{S}$, which features in the linear response formulation, does \textit{not} coincide with the thermal state of the final Hamiltonian $H(0)=H_0 +\delta \lambda\hat{N}_\mathcal{S}$ at the \textit{same initial temperature} $T$. In fact, this is exactly what happens in the two level system considered above. The work done by the time dependent Hamiltonian changes the internal energy of the system, and this is accounted for by a temperature change of order $\delta \lambda$.   

The general condition for the latter phenomena not to occur, is that the system under consideration is coupled to a thermal bath. In the examples studied in our work, we consider extended systems with a continuous gapless density of states. Hence the leads act as a thermal bath, and this infinite set of modes absorbs the energy change of order $\delta \lambda$ without causing a change in temperature of the bath.

Hence, the general condition for which $c(T) \ne 0$ is that the total specific heat capacity is finite. The metallic leads featuring in generic quantum impurity systems have a finite specific heat capacity per unit volume, meaning an infinite total specific heat capacity. Thus $c(T)=0$ in these systems.

%%%%%%%%%%%%%%%

\section{Derivation of Eq.~6}
In LR, the cumulant generating function $K(\eta) \equiv \ln \langle e^{-\eta \beta W_{diss}} \rangle$ admits an explicit expression \cite{guarnieri2023generalised}
\be
K(\eta)=-\int_0^\tau dt \dot{\lambda}(t) \int_0^\tau dt' \dot{\lambda}(t') \int \frac{d \omega}{2\pi} g_\eta(\omega) \tilde{\Psi}_0(\omega)e^{i\omega (t-t')},
\label{eq:Keta}
\ee
with $g_\eta(\omega) =\frac{\sinh(\frac{\beta\omega(1-\eta)}{2})\sin(\frac{\beta\omega\eta}{2})}{\beta\omega\sinh(\frac{\beta\omega}{2})}$. All the cumulants of the WDF are given by \cite{guarnieri2023generalised}
\begin{equation}\label{eq:cumulants_gf}
    2\pi 2\beta\kappa^n_W= 
    \begin{cases}
         \int d\omega\: \omega^{n-1}\:\tilde{\Psi}_0(\omega)\: \left|\int_0^{\tau}dt \dot{\lambda}(t)e^{i\omega t}\right|^2 & \text{: $n$ odd}\\
        \int d\omega\: \omega^{n-1} \:{\rm coth}(\tfrac{\omega}{2T})\: \tilde{\Psi}_0(\omega)\: \left|\int_0^{\tau}dt \dot{\lambda}(t)e^{i\omega t}\right|^2 & \text{: $n$ even}
    \end{cases}
\end{equation}
For linear ramp, $\lambda(t)=At/\tau$, thus we have $\left|\int_0^{\tau}dt \dot{\lambda}(t)e^{i\omega t}\right|^2=A^2{\rm{sinc}}^2(\omega\tau/2)$.
Substituting it into Eq.~\ref{eq:cumulants_gf}, and noticing Eq.~\ref{eq:Psi_fre} and Eq.~\ref{eq:SI_c}, we get
\begin{equation}\label{eq:cumulants_gf_more}
    \frac{2\pi \kappa^n_W}{A^2}= 
    \begin{cases}
%         \int d\omega\: \omega^{n-1}\: {\rm{sinc}}^2(\frac{\omega\tau}{2})\left(\frac{{\rm{Im}\chi(\omega)}}{\omega}-\pi(\frac{d\langle \hat{N}_{\mathcal{S}}\rangle_0}{d\lambda} + {\rm{Re}}\chi(\omega=0))\:\delta(\omega)\right) & \text{: $n$ odd}\\
%        \int d\omega\: \omega^{n-1} \:{\rm coth}(\tfrac{\omega}{2T})\: {\rm{sinc}}^2(\frac{\omega\tau}{2})\left(\frac{{\rm{Im}\chi(\omega)}}{\omega}-\pi(\frac{d\langle \hat{N}_{\mathcal{S}}\rangle_0}{d\lambda} + {\rm{Re}}\chi(\omega=0))\:\delta(\omega)\right) & \text{: $n$ even}
         \int d\omega\: \omega^{n-1}\: {\rm{sinc}}^2(\frac{\omega\tau}{2})\left(\frac{{\rm{Im}\chi(\omega)}}{\omega}-\pi\delta(\omega)\:\big[\frac{d\langle \hat{N}_{\mathcal{S}}\rangle}{d\lambda} + {\rm{Re}}\chi(\omega)\big]\right) & \text{: $n$ odd}\\
        \int d\omega\: \omega^{n-1} \:{\rm coth}(\tfrac{\omega}{2T})\: {\rm{sinc}}^2(\frac{\omega\tau}{2})\left(\frac{{\rm{Im}\chi(\omega)}}{\omega}-\pi\delta(\omega)\:\big[\frac{d\langle \hat{N}_{\mathcal{S}}\rangle}{d\lambda} + {\rm{Re}}\chi(\omega)\big]\right) & \text{: $n$ even}
    \end{cases}
\end{equation}
We note that the delta-function piece inside the integral contributes only for the $n=1$ and $n=2$ cumulants. For all $n\ge 3$ the cumulants are given exactly by Eq.~6 of the main text, even when $c(T)$ is finite. For $n=1$ and $n=2$ we pick up corrections, leading to:
\begin{equation}\label{eq:cumulants_gf_n1-2}
    \frac{2\pi\kappa^n_W}{A^2}= 
    \begin{cases}
         -\pi\left(\frac{d\langle \hat{N}_{\mathcal{S}}\rangle}{d\lambda} + {\rm{Re}}\chi(\omega=0)\right)+\int d\omega\: \omega^{n-2}\: {\rm{sinc}}^2(\frac{\omega\tau}{2})\:{\rm Im}\chi(\omega)& \text{: $n=1$}\\
        -2\pi T\left(\frac{d\langle \hat{N}_{\mathcal{S}}\rangle}{d\lambda} + {\rm{Re}}\chi(\omega=0)\right)+\int d\omega\: \omega^{n-2} \:{\rm coth}(\tfrac{\omega}{2T})\: {\rm{sinc}}^2(\frac{\omega\tau}{2})\:{\rm Im}\chi(\omega) & \text{: $n=2$}
    \end{cases}
\end{equation}

Finally, as we explained in Sec.\ref{se:ceqz}, we note that $c(T)=-\frac{d\langle \hat{N}_{\mathcal{S}}\rangle}{d\lambda}|_{\lambda=0} -{\rm{Re}}\chi(\omega=0)=0$ for the systems we consider. In this case, we get Eq.~6 of the main text for all $n$. Note also that in LR, up to order $A^2$, the cumulants and the moments of the dissipated work are equal.

%%%%%%%

\section{Derivation of Eqs.~9 and 10}
\label{se:9_10}
By definition, we have
\be
\langle \hat{N}_{\mathcal{S}} \rangle_{\lambda_0+\delta\lambda} = \frac{{\rm{Tr}}\left[\hat{N}_{\mathcal{S}}e^{-\beta(\hat{H}_0 + \delta\lambda\hat{N}_{\mathcal{S}})}\right]}{{\rm{Tr}}e^{-\beta(\hat{H}_0 + \delta\lambda\hat{N}_{\mathcal{S}})}}. 
\label{eq:ns_up_change}
\ee
Using the identity
\be
e^{\hat{A}+\alpha \hat{B}} = e^{\hat{A}} + \alpha\int_0^1 dy e^{y\hat{A}} \hat{B} e^{(1-y)\hat{A}} + \mathcal{O}(\alpha^2), 
\ee
we could expand both the numerator and denominator of Eq.~\ref{eq:ns_up_change} to linear- order in $\delta\lambda$, which gives
\begin{align*}
    {\rm{Tr}}\left[\hat{N}_{\mathcal{S}}e^{-\beta(\hat{H}_0 + \delta\lambda\hat{N}_{\mathcal{S}})}\right]&={\rm{Tr}}\left[\hat{N}_{\mathcal{S}}e^{-\beta \hat{H}_0}\right]-\beta\delta\lambda\int_0^1dy{\rm{Tr}}\left[\hat{N}_{\mathcal{S}}e^{-\beta y\hat{H}_0}\hat{N}_{\mathcal{S}}e^{-\beta(1-y)\hat{H}_0}\right],\\
    {\rm{Tr}}e^{-\beta(\hat{H}_0 + \delta\lambda\hat{N}_{\mathcal{S}})}&={\rm{Tr}}e^{-\beta \hat{H}_0}-\beta\delta\lambda{\rm{Tr}}\left[\hat{N}_{\mathcal{S}}e^{-\beta \hat{H}_0}\right].
\end{align*}
Thus, to linear-order in $\delta\lambda$, we have
\be
\langle \hat{N}_{\mathcal{S}} \rangle_{\lambda_0+\delta\lambda}
=\langle \hat{N}_{\mathcal{S}} \rangle_{0} + \beta\delta\lambda\langle \hat{N}_{\mathcal{S}} \rangle_{0}^2-\delta\lambda\int_0^{\beta}ds \langle \hat{N}_{\mathcal{S}}(-is) \hat{N}_{\mathcal{S}}(0) \rangle_0,
\ee
and Eq.~9 directly follows.

To derive Eq.~10, we start from  Eq.~\ref{eq:Psi_dt}. Considering $\frac{1}{\beta}\frac{d^2\Psi_0(t)}{dt^2}|_{t=0}$, we find
\begin{align}
  \left.\frac{1}{\beta}\frac{d^2\Psi_0(t)}{dt^2}\right\vert_{t=0}&=\left.-i\frac{d}{dt}{\rm{Tr}}\left[{\rho_0 [\hat{N}_{\mathcal{S}}(t), \hat{N}_{\mathcal{S}}]}\right]\right\vert_{t=0}\nonumber\\
  &=\left.-i\frac{d}{dt}\left({\rm{Tr}}\left[\rho_0 e^{i\hat{H}_0 t}\hat{N}_{\mathcal{S}}e^{-i\hat{H}_0 t}\hat{N}_{\mathcal{S}}\right] - {\rm{Tr}}\left[\rho_0 \hat{N}_{\mathcal{S}}e^{i\hat{H}_0 t}\hat{N}_{\mathcal{S}}e^{-i\hat{H}_0 t}\right]\right)\right\vert_{t=0}\nonumber\\
  &=2\left({\rm{Tr}}\left[\rho_0\hat{H}_0\hat{N}_{\mathcal{S}}^2\right] - {\rm{Tr}}\left[\rho_0\hat{N}_{\mathcal{S}}\hat{H}_0\hat{N}_{\mathcal{S}}\right]\right)\nonumber\\
  &=-{\rm{Tr}}\left[\rho_0 \left[\hat{N}_{\mathcal{S}}, [\hat{H}_0, \hat{N}_{\mathcal{S}}]\right]\right].
\end{align}
Notice that $\frac{d^2\Psi_0(t)}{dt^2}|_{t=0}=-\int d\omega \omega^2\tilde{\Psi}_0(\omega)$, thus
\be
\frac{1}{\beta}\int d\omega \omega^2\tilde{\Psi}_0(\omega) = {\rm{Tr}}\left[\rho \left[\hat{N}_{\mathcal{S}}, [\hat{H}_0, \hat{N}_{\mathcal{S}}]\right]\right].
\label{eq:kappa3_sudden}
\ee

Finally, notice that in our setup, $[\hat{H}_0, \hat{N}_{\mathcal{S}}]=[\hat{H}_{\mathcal{SE}}, \hat{N}_{\mathcal{S}}]$. In addition, the dot-lead hybridization term $\hat{H}_{\mathcal{SE}}$ can be decomposed into a tunneling-in part and a tunneling-out part, $\hat{H}_{\mathcal{SE}}=\hat{H}_{\mathcal{SE}}^{(+)} + \hat{H}_{\mathcal{SE}}^{(-)}$, where we have $[\hat{N}_{\mathcal{S}}, \hat{H}_{\mathcal{SE}}^{(+)}]=q\hat{H}_{\mathcal{SE}}^{(+)}$ and $[\hat{N}_{\mathcal{S}}, \hat{H}_{\mathcal{SE}}^{(-)}]=-q\hat{H}_{\mathcal{SE}}^{(-)}$ for the tunneling process that involves $q$ charges. For both the resonant level model and the charge Kondo device, $q=1$, thus we have 
\be
\left[\hat{N}_{\mathcal{S}}, [\hat{H}_0, \hat{N}_{\mathcal{S}}]\right]=-\left[\hat{N}_{\mathcal{S}}, [ \hat{N}_{\mathcal{S}}, \hat{H}_0]\right]=-\left[\hat{N}_{\mathcal{S}}, [ \hat{N}_{\mathcal{S}}, \hat{H}_{\mathcal{SE}}^{(+)} + \hat{H}_{\mathcal{SE}}^{(-)}]\right]=-\left[\hat{N}_{\mathcal{S}},  \hat{H}_{\mathcal{SE}}^{(+)} - \hat{H}_{\mathcal{SE}}^{(-)}\right]=-\left(\hat{H}_{\mathcal{SE}}^{(+)} + \hat{H}_{\mathcal{SE}}^{(-)}\right)=-\hat{H}_{\mathcal{SE}}.
\label{eq:hse_equiv}
\ee

Combining Eq.~\ref{eq:kappa3_sudden} with Eq.~\ref{eq:hse_equiv}, we get Eq.~10.
%###################
%###################

\section{Resonant level model}
\begin{figure}[h]
    \centering
    \includegraphics[width=0.7\linewidth]{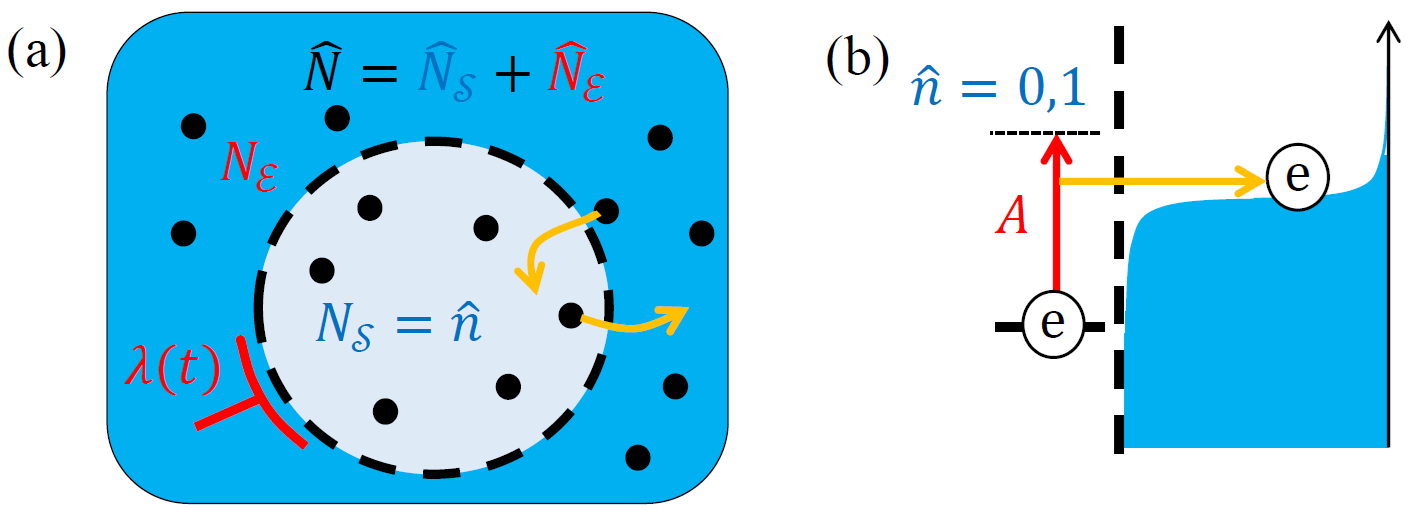}
    \caption{(a) Generic scenario in which a globally-conserved charge $\hat{N}=\hat{N}_{\mathcal{S}}+\hat{N}_{\mathcal{E}}$ is shared by the system and the environment. (b) In a QD, where $\hat{N}_{\mathcal{S}}\equiv \hat{n}$ corresponds to the dot occupation, driving the dot potential $\epsilon_d$ induces nontrivial dynamics that can be characterized by quantum work statistics. }
    \label{fig:rlm}
\end{figure}

A noninteracting QD is described by the spinless RLM. The `system' is the QD level itself, while the `environment' is the bath of conduction electrons, see Fig.~\ref{fig:rlm}(a). They are coupled by the dot-lead hybridization term $\hat{H}_{\mathcal{SE}}$ which allows the dot charge to fluctuate, even though the total particle number is fixed. Fig.~\ref{fig:rlm}(b) illustrates the QD charge dynamics induced by driving the local dot potential. Here we treat the environment explicitly, providing exact results beyond the Markovian approximation used in the master equation, which neglects system-environment entanglement.

Explicitly, we have
\begin{subequations}
\begin{align}
\langle \hat{N}_{\mathcal{S}} \rangle_0 &= \frac{1}{2} + \frac{1}{2\pi}{\rm{Im}}\left(\psi(\frac{1}{2} + \frac{\Gamma - i\epsilon_d}{2\pi T}) - \psi(\frac{1}{2} + \frac{\Gamma + i\epsilon_d}{2\pi T})\right),\\
\langle \hat{H}_{\mathcal{SE}} \rangle_0 &= 2 \Gamma \int_{-\Lambda}^\Lambda \frac{d \omega}{\pi} f(\omega)\frac{\omega-\epsilon_d}{(\omega-\epsilon_d)^2+\Gamma^2},\\
\chi(\omega) =& -\frac{1}{\pi\Gamma}\int_{-\infty}^{\infty} dx f(\Gamma x + \epsilon_d)\frac{2x}{(x^2+1)\left(x^2 - (\omega/\Gamma + i)^2\right)},\label{eq:rlm_chi}
\end{align}
\end{subequations}
where $\psi(x)$ is the digamma function. Eq.~\ref{eq:rlm_chi} was previously derived in Ref.~\onlinecite{filippone:tel-00908428}, see Eq.(1.55) thereof. At the particle-hole symmetric point $\epsilon_d=0$, the integral in Eq.\ref{eq:rlm_chi} can be done and we find a closed-form expression for ${\rm{Im}}\chi(\omega)$, 

\be
{\rm{Im}}\chi(\omega) = {\rm{Im}}\left[\frac{2}{\pi\Gamma}\frac{1}{\frac{\omega^2}{\Gamma^2}+2i\frac{\omega}{\Gamma}}\times\right. \\
  \left.\left(\psi(\frac{1}{2} + \frac{\Gamma}{2\pi T}) - \psi(\frac{1}{2} + \frac{\Gamma}{2\pi T} - i\frac{\omega}{2\pi T})\right)\right].  
\ee

\bibliography{bibliography}